\documentclass[aps,prd,twocolumn,superscriptaddress,nofootinbib]{revtex4-1}
\usepackage{amsmath}
\usepackage[scientific-notation=true]{siunitx}
\usepackage{graphicx}
\usepackage{hyperref}
\usepackage{placeins}
\usepackage[english]{babel}
\let\vec\mathbf
\global\hyphenpenalty=450
\global\interfootnotelinepenalty=10000

\begin{document}
\title{Density profiles of ultracompact minihalos: Implications for constraining the primordial power spectrum}
\author{M. Sten Delos}
\email{Electronic address: delos@unc.edu}
\author{Adrienne L. Erickcek}
\email{Electronic address: erickcek@physics.unc.edu}
\affiliation{Department of Physics and Astronomy, University of North Carolina at Chapel Hill, Phillips Hall CB3255, Chapel Hill, North Carolina 27599, USA}
\author{Avery P. Bailey}
\affiliation{Department of Astrophysical Sciences, Princeton University, Peyton Hall,
	Princeton, New Jersey 08544, USA}
\affiliation{Department of Physics and Astronomy, University of North Carolina at Chapel Hill, Phillips Hall CB3255, Chapel Hill, North Carolina 27599, USA}
\author{Marcelo A. Alvarez}
\affiliation{Berkeley Center for Cosmological Physics, Campbell Hall 341, University of California, Berkeley, California 94720, USA}

\begin{abstract}
Enhanced density fluctuations on small scales would lead to the formation of numerous dark matter minihalos, so limits on the minihalo abundance can place upper bounds on the small-scale primordial power spectrum.  In particular, the ultracompact minihalo (UCMH), a dark matter structure hypothesized to possess a $\rho\propto r^{-9/4}$ density profile due to its formation at $z\geq 1000$, has been used to establish an upper bound on the primordial power spectrum at scales smaller than 2 Mpc.  The extreme slope of this density profile amplifies the observational signals of UCMHs.  However, we recently showed via N-body simulations that the $\rho\propto r^{-9/4}$ density profile does not develop in realistic formation scenarios, throwing UCMH-derived power spectrum constraints into question.  Instead, minihalos develop shallower inner profiles with power-law indices between $-3/2$ and $-1$.  In this paper, we expand on that result and discuss its implications.  Using a model that is calibrated to simulation results and predicts halo structures in spiked power spectra based on their formation times, we calculate new upper bounds on the primordial power spectrum based on limits on the dark matter annihilation rate within the Galaxy.  We find that despite assuming shallower profiles, this minihalo model actually yields stronger constraints than the previous UCMH picture owing to its inclusion of all minihalos instead of only the earliest-forming ones.
\end{abstract}

                              
\maketitle

\section{Introduction}

Ultracompact dark matter minihalos have emerged as a powerful probe of early-Universe physics.  Overdense regions with $\delta\equiv \delta\rho/\rho \gtrsim 10^{-3}$ at horizon entry seed the formation of dark matter minihalos near the time of recombination ($z\simeq 1000$) \cite{ricotti2009new}, and such early formation yields highly compact structures potentially visible through dark matter annihilation \cite{scott2009gamma,saito2011primordial,yang2011constraints,berezinsky2013formation,1712.01724,nakama2018constraints,josan2010gamma,bringmann2012improved,yang2013neutrino,yang2017tau,yang2011abundance,zhang2011impact,yang2016contributions,clark2017heating,yang2011new,yang2013contribution,beck2018through} or by their gravitational signatures \cite{ricotti2009new,zackrisson2013hunting,clark2015investigatingI,*clark2016erratumI,li2012new}.  The nondetection of these structures thus constrains the amplitude of primordial density fluctuations, making it a probe of the primordial power spectrum \cite{josan2010gamma,bringmann2012improved,li2012new,yang2013neutrino,yang2013dark,yang2014constraints,clark2015investigatingII,*clark2017erratumII,yang2017tau,nakama2018constraints} and hence of inflationary models \cite{aslanyan2016ultracompact} and the thermal history of the Universe \cite{choi2017new}.

These ultracompact minihalos (UCMHs) provide access to perturbations on scales too small to be directly observed.  Cosmic microwave background (CMB) observations indicate that the primordial power spectrum of curvature fluctuations $\mathcal{P}_\zeta(k)$ is consistent with a slightly red-tilted but otherwise featureless power law \cite{hlozek2012atacama} with amplitude $\mathcal{A}_s = (2.142\pm 0.049)\times10^{-9}$ \cite{2016planck}, and the Lyman-$\alpha$ forest tells a similar story \cite{bird2011minimally}.  However, these observations are only able to probe wavelengths longer than \SI{2}{Mpc}, and numerous inflationary models predict an enhancement in small-scale power
\cite{joy2008new,salopek1989designing,starobinsky1992,ivanov1994inflation,starobinsky1998beyond,chung2000probing,barnaby2009particle,barnaby2010features,silk1987double,polarski1992spectra,adams1997multiple,achucarro2011features,cespedes2012importance,randall1996supernatural,stewart1997flattening,copeland1998black,covi1999running,covi1999observational,martin2000nonvacuum,martin2001trans,ben2010cosmic,gong2011waterfall,lyth2011contribution,bugaev2011curvature,barnaby2011large,barnaby2012gauge}.
Certain nonstandard thermal histories, such as an early matter-dominated era \cite{erickcek2011reheating,barenboim2014structure,fan2014nonthermal,erickcek2015dark} or an era dominated by a fast-rolling scalar field \cite{redmond2018growth}, also enhance small-scale fluctuations.  Thus, probing the small-scale power spectrum is key to understanding early-Universe physics.

Unfortunately, at sub-Mpc scales, we only have upper bounds on density fluctuations, which are obtained through the absence of secondary effects.  Density contrasts of order $0.3$ at horizon entry would form primordial black holes, so constraints on their abundance constrain ${\mathcal P_\zeta(k) \lesssim \num{3e-2}}$ over a wide range of scales \cite{josan2009generalized}.  An excess of integrated power would imprint distortions onto the CMB blackbody spectrum, so their nonobservation constrains $\mathcal P_\zeta(k) \lesssim \num{2e-5}$ for $k \lesssim 10^4\ \mathrm{Mpc}^{-1}$ \cite{chluba2012probing}.  However, UCMHs supply the strongest constraints.  The nondetection of gamma rays from dark matter annihilation in UCMHs constrains $\mathcal P_\zeta(k) \lesssim \num{3e-7}$ for ${k \lesssim 10^7\ \mathrm{Mpc}^{-1}}$ \cite{bringmann2012improved}.

However, with one recent exception \cite{nakama2018constraints}, all constraints derived from UCMHs have been calculated assuming they develop the $\rho\propto r^{-9/4}$ density profile, which is drawn from analytic radial infall theory \cite{fillmore1984self,bertschinger1985self} and taken to apply to halos forming at $z\gtrsim 1000$ due to the small velocity dispersion at those times \cite{ricotti2009new}.  This profile has a much steeper inner form than is typically observed in simulations (e.g., \cite{frenk2012dark}), a property that enhances the observational signatures of UCMHs.  The applicability of this profile was first called into question in Ref.~\cite{gosenca20173d}, and in Ref.~\cite{Delos2018ultracompact}, hereafter Paper~I, we showed by means of N-body simulations that halos forming in a Gaussian field---even UCMHs forming as early as $z=1000$ from fluctuations as extreme as $6.8\sigma$---do not develop the $\rho\propto r^{-9/4}$ density profile.  Instead, they develop shallower inner density profiles: $\rho\propto r^{-\gamma}$ with $1\leq \gamma\leq 3/2$.  In this paper, we present our results in greater detail and discuss the implications of this discovery.  In addition to the $\rho\propto r^{-9/4}$ assumption, previous UCMH-derived power spectrum constraints employed only the minihalos that form at $z\geq 1000$.  Since we have shown that all minihalos develop shallower density profiles, there is no need to make this restriction.  We show that the resulting new bounds on the power spectrum are stronger than the previous UCMH constraints.

The $\rho\propto r^{-9/4}$ density profile has been taken to be a consequence of nearly radial mass infall onto a halo that formed at $z\gtrsim 1000$ \cite{ricotti2009new}, so we replicate this scenario as closely as possible in our simulations by finding extremely rare $6.8\sigma$ density peaks that collapse near $z=1000$.  The UCMH formation scenario is tested in two power spectra at the opposite extremes that are motivated by inflationary phenomenology.  First, we use a spiked power spectrum with fluctuations enhanced over a narrow range of scales.  Second, we use a stepped power spectrum with fluctuations enhanced over all scales accessible to the simulation.  In the narrowly enhanced power spectrum, halos develop in relative isolation, a situation that might be expected to reproduce the radial infall solution.  However, we find that all halos, even the UCMHs forming at $z\simeq 1000$, develop $\rho\propto r^{-3/2}$ inner density profiles.  In fact, this profile also appears in another context: it is the density profile seen in the smallest halos forming above a free-streaming cutoff \cite{ishiyama2010gamma,anderhalden2013density,*anderhalden2013erratum,ishiyama2014hierarchical,polisensky2015fingerprints,angulo2017earth,ogiya2017sets}.  Meanwhile, the broadly enhanced power spectrum builds halos hierarchically from smaller halos and yields density profiles of the Navarro-Frenk-White (NFW) form \cite{navarro1995simulations,navarro1996structure,navarro1997universal},
\begin{equation}\label{NFW}
\rho(r) = \frac{\rho_s}{(r/r_s)(1+r/r_s)^2},
\end{equation}
with a $\rho\propto r^{-1}$ inner profile.  This is the same form that appears in simulations of galaxy-scale structure.  Evidently, UCMHs, which we define as halos forming at $z\geq 1000$, develop the same density profiles as halos that form at much later times.

We also introduce a new model for predicting the density profiles of minihalos that form from spiked power spectra based on their formation times.  Spectral spikes can arise from steps in the inflaton potential \cite{salopek1989designing,starobinsky1992,ivanov1994inflation,starobinsky1998beyond} or from particle production during inflation \cite{chung2000probing,barnaby2009particle,barnaby2010features}.  Near the free-streaming cutoff, the power spectrum imprinted by an early matter-dominated era is also similar to a spike \cite{erickcek2011reheating,barenboim2014structure,fan2014nonthermal,erickcek2015dark}.  Moreover, spiked power spectra are less well constrained than flatter power spectra by CMB spectral distortions, which limit the power integrated over a broad range in $k$-space \cite{chluba2012probing}.  In this paper, we begin an investigation of halos forming from spiked power spectra that we will expand upon in the next paper of this series \cite{delos2018predict}, hereafter Paper~III (in preparation).

Finally, we discuss the impact of our result on the capacity for minihalos to constrain the primordial power spectrum.  We use our model to calculate an upper bound on the amplitude of spiked power spectra that incorporates the new shallower minihalo density profiles.  This upper bound is based on limits from Fermi-LAT \cite{atwood2009large} on gamma rays from dark matter annihilation.  Despite the reduced annihilation rate implied by the shallower profile, this constraint is stronger than an equivalent UCMH constraint derived using the $\rho\propto r^{-9/4}$ density profile.  Our model provides a stronger constraint because it accounts for all halos, whereas the old UCMH model only counted halos forming at $z\gtrsim 1000$.  Our calculation demonstrates the continued viability of minihalos as probes of the small-scale power spectrum, and we discuss future avenues for improvement.

This paper is organized as follows.  In Sec.~\ref{sec:setup}, we select power spectra and detail the setup of our simulations.  Section~\ref{sec:ucmh} presents the simulation results.  The UCMH density profile is the main result, but we also make remarks on UCMH growth and the effects of mergers.  In Sec.~\ref{sec:later}, we sample later-forming minihalos to develop a general model for minihalo density profiles, and Sec.~\ref{sec:constraints} employs this model to calculate new constraints on the primordial power spectrum.  In Sec.~\ref{sec:constraint_discussion}, we discuss how the new minihalo picture differs from the old UCMH picture.  Section~\ref{sec:conclusion} concludes and outlines the ways in which our calculation can be improved in future work.  Appendices \ref{sec:rad} and~\ref{sec:conv} contain additional information about our simulations, including numerical convergence tests.  Appendices \ref{sec:ptunif}--\ref{sec:bsa} contain further details about our calculation of the power spectrum constraint.

\section{Simulation preparation}\label{sec:setup}

We carry out simulations of halos forming at ${z\simeq 1000}$ from extreme peaks in the density field.  This picture is intended to match the UCMH formation scenario \cite{ricotti2009new}, and we aim to show conclusively that the ${\rho\propto r^{-9/4}}$ single-power-law density profile does not arise in halos forming due to an enhancement of the primordial power spectrum.

\subsection{Power spectrum}

In order to perform numerical experiments on such early-forming minihalos, we must start with an enhanced power spectrum.  Inflationary models supply a rich phenomenology in this respect.  Steps, kinks, or second-derivative jumps in the inflaton potential would imprint spikes, steps, or bends, respectively, on the primordial power spectrum
\cite{salopek1989designing,starobinsky1992,ivanov1994inflation,starobinsky1998beyond,joy2008new}.
Particle production during inflation can produce a spike in the power spectrum
\cite{chung2000probing,barnaby2009particle,barnaby2010features},
while multifield inflation can imprint steps
\cite{silk1987double,polarski1992spectra,adams1997multiple}
or oscillations
\cite{achucarro2011features,cespedes2012importance}.  Inflation aside, an early matter-dominated era enhances perturbations that enter the horizon prior to the onset of radiation domination \cite{erickcek2011reheating,barenboim2014structure,fan2014nonthermal,erickcek2015dark}, and an era dominated by a fast-rolling scalar field generates a similar enhancement \cite{redmond2018growth}.

\begin{figure}[t]
	\centering
	\includegraphics[width=\columnwidth]{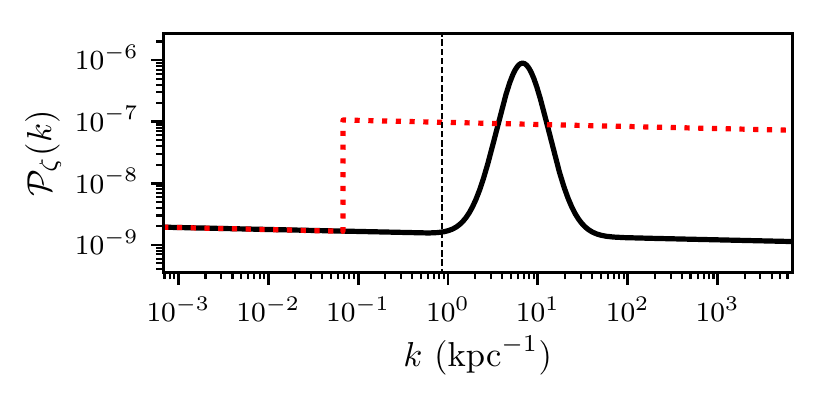}
	\caption{\label{fig:prim} The dimensionless primordial power spectrum of curvature fluctuations used in our UCMH simulations.  The solid line shows the spike modification, while the dotted line shows the step.  The vertical dashed line indicates the smallest $k$ (largest scale) accessible in the simulations.}
\end{figure}

For our simulations, we consider two examples from these possible power spectrum enhancements.  First, we consider a narrow spike in the power spectrum.  This shape has possible inflationary origins, as discussed above, and is also qualitatively similar to the enhancement generated by an early matter-dominated era at scales close to the free-streaming cutoff.  Next, we consider a step in the power spectrum, intended to represent the opposite extreme where fluctuations are enhanced over a broad range of scales.  The two power spectra are plotted in Fig.~\ref{fig:prim}.  We superpose these modifications on a conventional power spectrum with amplitude $\mathcal{A}_s = \num{2.142e-9}$ and spectral index $n_s = 0.9667$ \cite{2016planck}.  The spike contains 90\% of its added power inside 1 e-fold in $k$, while the step amplifies fluctuations over the full range of scales accessible to the simulation.  We will focus on the spiked power spectrum for most of Sec.~\ref{sec:ucmh} and return to the step in Sec.~\ref{sec:step}.

Halos forming from more extreme density contrasts are both more spherically symmetric \cite{bardeen1986statistics} and less affected by nearby structure.  To best simulate the UCMH formation scenario, we tune our power spectra so that halos forming by $z=1000$ are exceedingly rare.  In particular, the spiked power spectrum is tuned so that a $6.8\sigma$ fluctuation is necessary to seed such early collapse, and we generate a large number of random fields in order to obtain a handful of boxes to use as initial conditions for our simulations.  This procedure may be contrasted with that of Ref.~\cite{gosenca20173d}, who simulated a typical box whose most extreme peak was $4.3\sigma$.  UCMHs forming from peaks as extreme as $6\sigma$ are employed to derive observational constraints \cite{bringmann2012improved}, so we wish to exceed this amplitude to conclusively rule out the $\rho\propto r^{-9/4}$ profile.

\subsection{Simulation setup}\label{sec:IC}

A matter power spectrum is calculated at $z=1000$ from the primordial power spectrum using the Boltzmann code \textsc{Camb Sources} \cite{challinor2011linear,lewis200721}.  To match simulation behavior, this power spectrum is evolved back to an earlier time using the M\'esz\'aros equation \cite{meszaros1974behaviour}
\begin{equation}\label{meszaros}
\frac{d^2\delta}{dy^2}+\frac{2+3y}{2y(y+1)}\frac{d\delta}{dy}-\frac{3}{2y(y+1)}\delta=0,
\end{equation}
which describes the subhorizon (Newtonian) evolution of dark matter density perturbations when baryons and radiation fluctuations are neglected.  Here $y\equiv a/a_\mathrm{eq}$, where $a_\mathrm{eq}$ is the scale factor at matter-radiation equality.  The physical solution to this equation is obtained by matching its general solution to the asymptotic behavior $\delta\propto\ln(0.44a/a_H)$ during radiation domination, where $a_H$ is the scale factor when the perturbation mode enters the horizon.  This physical solution is \cite{hu1996small}
\begin{align}\label{evo}
\delta\propto
&\left[\ln\left(\frac{k}{0.12 h\ \si{Mpc^{-1}}}\right)
-\ln\left(\frac{\sqrt{1+y}+1}{\sqrt{1+y}-1}\right)
\right]\left(y+\frac{2}{3}\right)\nonumber\\
&+ 2\sqrt{1+y},
\end{align}
which provides a convenient prescription for calculating the evolution of a density contrast $\delta$ at linear order during mixed matter-radiation domination.

We choose to study fluctuations of order 0.2 kpc, so the spectral spike of Fig.~\ref{fig:prim} is centered at wave number $k_s=\SI{6.8}{kpc^{-1}}$.  The starting redshift is chosen to be $z=8\times10^6$ so that a density contrast that collapses at $z\simeq 1000$ is initially of order $0.1$.  We do not expect our results to depend significantly on either of these choices.  We fix a comoving box size of 7.4 kpc and search periodic Gaussian random fields generated at the initial redshift for candidate peaks to collapse near $z=1000$.  Our search proceeds by first generating a Gaussian random field on a grid at the initial redshift using our spiked power spectrum.  We then linearly evolve that field to $z=1000$ and check whether the evolved density field has a peak\footnote{For simplicity, we require $\delta>1.686$ in one grid-cell in our $512^3$-cell density field, which corresponds to a smoothing scale of \SI{0.014}{kpc}.  Because the power is concentrated in the spike, the precise choice of smoothing scale is unimportant.} with $\delta>1.686$, the linear threshold for collapse.  If so, we use that grid, and if not, we generate a new one.  Once we have a suitable density field, we use the Zel'dovich approximation to perturb a particle grid into a corresponding initial particle distribution.  Since our simulations begin while the Universe is radiation-dominated, initial velocities are computed by differentiating Eq.~(\ref{evo}); see Appendix \ref{sec:rad} for details.

\begin{figure}[t]
  \centering
    \includegraphics[width=.8\columnwidth]{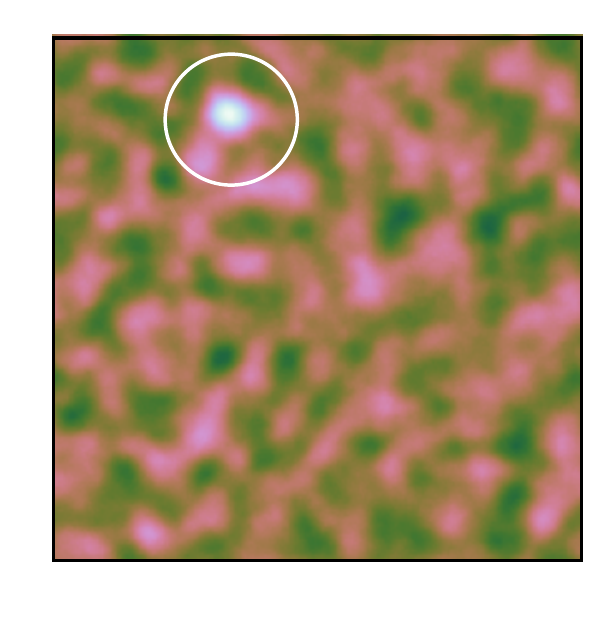}
    \caption{\label{fig:init} A slice of the 7.4 kpc density field used as initial conditions for the primary simulation run.  Lighter regions are denser.  The circle indicates the spherical region for a high-particle-density simulation.}
\end{figure}

For the spiked power spectrum shown in Fig.~\ref{fig:prim}, we generate 2.3 million random density fields.  Nine of them meet the collapse criterion, so we use these as the initial density fields and simulate them to $z=50$.  We also pick out one such density field, which we label the primary, to simulate at higher resolution and perform convergence tests; a slice of its initial density field is shown in Fig.~\ref{fig:init}.  Notice how extreme the most overdense region is compared to its surroundings: this is indeed a rare event.

We also resimulate each of these density fields with increased simulation-particle density by resampling the initial field at higher resolution and including only a sphere of radius 0.93 kpc (with vacuum boundary conditions) around the most overdense point.  This cut-out region is drawn in Fig.~\ref{fig:init}.  This procedure allows us to probe smaller scales, and in Appendix \ref{sec:conv}, we demonstrate that it does not change the density profile of the UCMH in the primary density field at $z=100$.  This convergence does not hold for the UCMHs in all nine fields: some of them begin to be influenced by structure outside the sphere as early as $z\sim 200$.  Consequently, we only carry out these cut-out simulations up to $z=400$ for the other eight density fields.

\subsection{N-body code}

We use the cosmological simulation code \textsc{Gadget-2} \cite{springel2005cosmological,springel2001gadget} for our numerical experiments.  \textsc{Gadget-2} is a hybrid N-body code that computes short-range forces using a tree method and long-range forces using Fourier techniques on a mesh.  A discussion on our choices of simulation parameters can be found in Appendix \ref{sec:conv}, along with convergence studies.  We also model all matter as collisionless dark matter with $\Omega_m=0.3089$ \cite{2016planck}: at the scales we study, dark matter halos cannot capture significant baryon content.

In order to accurately simulate a halo collapse at $z\simeq 1000$, our experiments must begin during radiation domination, so our N-body code must account for radiation.  However, fluctuations in the radiation density field decay rapidly after horizon entry (see e.g. \cite{dodelson2003modern}), so it is only necessary to model the effect of a smooth radiation component  on the expansion rate.  We modified the publicly available release of \textsc{Gadget-2} to include such a radiation component.  Tests of the accuracy of this code can be found in Appendix \ref{sec:rad}.

\section{Simulation results}\label{sec:ucmh}

\begin{figure}[t]
  \centering
    \includegraphics[width=\columnwidth]{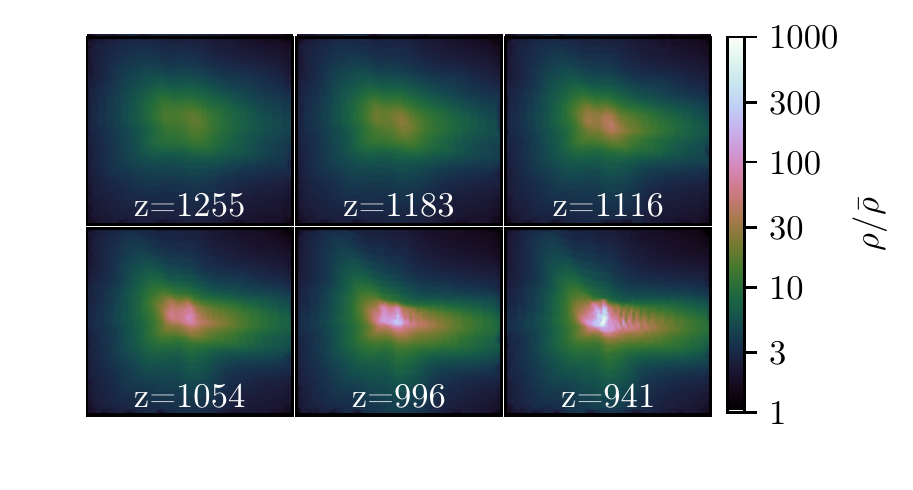}\\
    \vspace{-6mm}
    \includegraphics[width=.9\columnwidth]{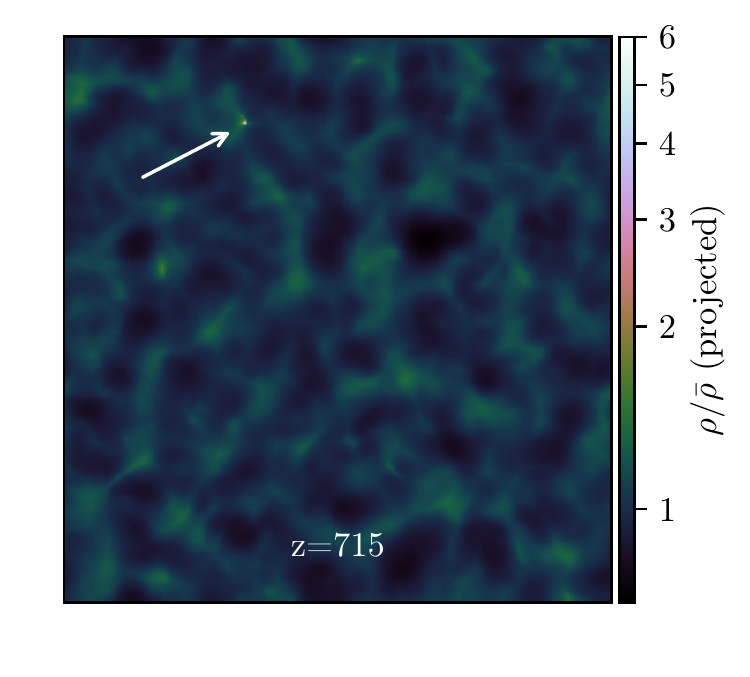}
    \caption{\label{fig:collapse} The density field for the primary run at different redshifts.  Top: A $(0.24\  \si{kpc})^2\times 0.06\ \si{kpc}$ slice showing the collapse of the UCMH near $z=1000$.  The color scale is logarithmic in units of the background matter density.  Bottom: The full $(\SI{7.4}{kpc})^3$ projected density field at $z=715$.  There is still only one halo, a testament to its rarity.}
\end{figure}

A visual inspection of the primary simulation box yields some key insights.  First, we note that our criterion for early collapse, that the linear density contrast be $\delta>1.686$ by $z=1000$, has worked as expected.  Figure~\ref{fig:collapse} shows a slice of the density field evolving from $z=1255$ to $z=941$ at the location of the extreme density peak where we expect the UCMH to form, and we see that the density at the central point grows astronomically around $z=1000$, an indication of collapse.  To emphasize the rarity of this event, we also show the density field at $z=715$: the UCMH is still the only halo to have collapsed by this redshift.

\begin{figure}[t]
  \centering
    \includegraphics[width=\columnwidth]{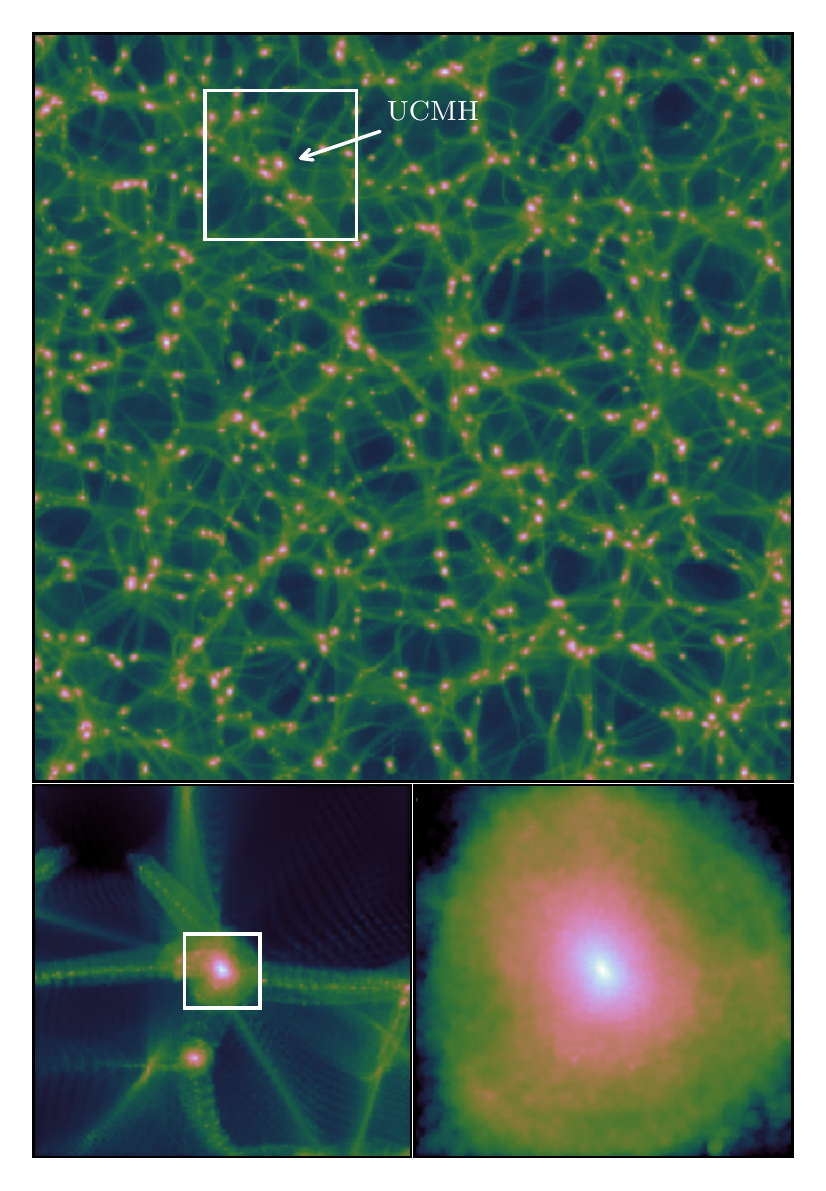}
    \caption{\label{fig:halo} The projected density field of the primary simulation box at $z=100$.  Top: The full 7.4 kpc field.  Bottom: Expanded pictures of the UCMH.  The left (right) panel shows the projected density field for the surrounding 1.5 kpc (0.3 kpc) cube.  Note that the expanded pictures do not fully match the white boxes because they are projected over smaller depths.}
\end{figure}

Next, we look at the density field at a much later redshift.  Figure~\ref{fig:halo} shows the density field at $z=100$ projected along one axis.  The imprint of the spike in the power spectrum is evident, for we see an almost uniform distribution of halos with no large-scale structure.  This is quite unlike a hierarchical growth picture (cf. Fig.~\ref{fig:step}).  There is also minimal small-scale structure: these halos appear generally isolated and are only linked by filaments.  These points are emphasized in the enhanced pictures of the main halo, where we see more clearly the lack of small-scale structure.  We also see the beginning of fragmentation of the filaments into halos, but this fragmentation is a numerical artifact; see Appendix~\ref{sec:conv}.

According to the \textsc{Rockstar} halo finder \cite{behroozi2012rockstar}, there are 530 halos with masses above $1.5 M_\odot$ at $z=100$, and these halos contain 24\% of the total mass of the simulation box within their virial radii.  Such an abundance of halos is clearly expected in any picture that can produce a halo that collapses by $z\simeq 1000$, but later halos have been neglected in prior UCMH treatments because they are expected to be less compact.  We will explore in Sec.~\ref{sec:later} whether younger halos have the same structure as the oldest ones.

\subsection{Density profiles}\label{sec:density_profiles}

\begin{figure*}[t]
  \centering
    \includegraphics[width=\textwidth]{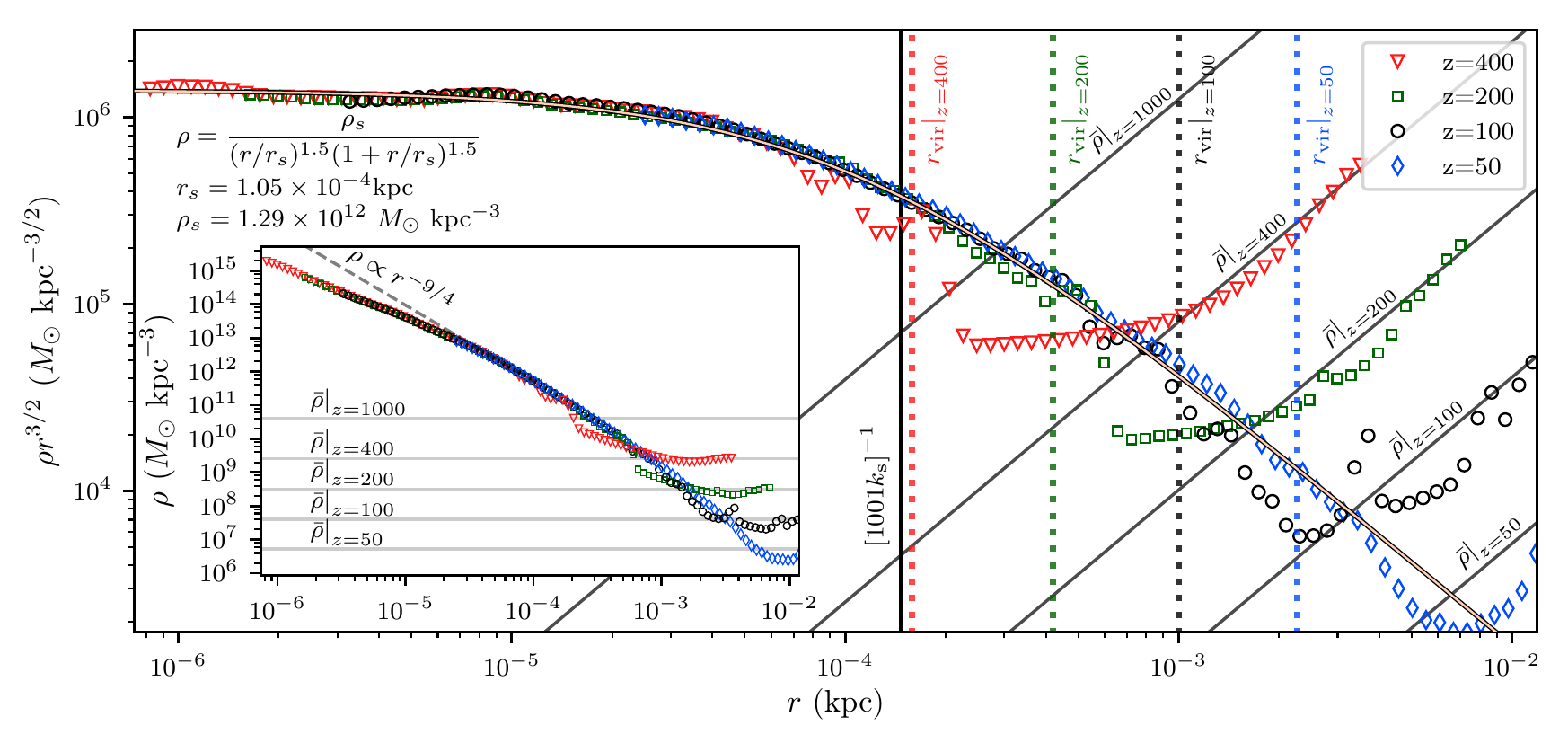}
    \caption{\label{fig:ucmh} The spherically averaged density profile of the UCMH in the primary density field at $z=400$, $z=200$, $z=100$, and $z=50$.  The vertical axis is scaled by $r^{3/2}$ to reduce the vertical range and better exhibit asymptotic behaviors; this practice will be adopted without remark in later figures.  The density profile approaches $\rho\propto r^{-3/2}$ at small $r$ and $\rho\propto r^{-3}$ at large $r$ and is fit well by Eq.~(\ref{Moore}) (solid curve).  The solid vertical line shows the physical scale of the power spectrum spike at $z=1000$, while the vertical dashed lines show the halo virial radius at different redshifts.  Inset: The same plot without y-axis scaling.  A $\rho\propto r^{-9/4}$ curve is shown for comparison.  We plot physical, not comoving, quantities.  We show results from the vacuum-bounded sphere inside $r_\mathrm{vir}$ for $z\leq 100$, and results from the full box otherwise.  The smallest radius at each redshift is set by $r>2.8\epsilon$, where $\epsilon$ is the force-softening length parameter (see Appendix~\ref{sec:conv}), and contains $N>3000$ particles.}
\end{figure*}

\begin{figure}[t]
	\centering
	\includegraphics[width=\columnwidth]{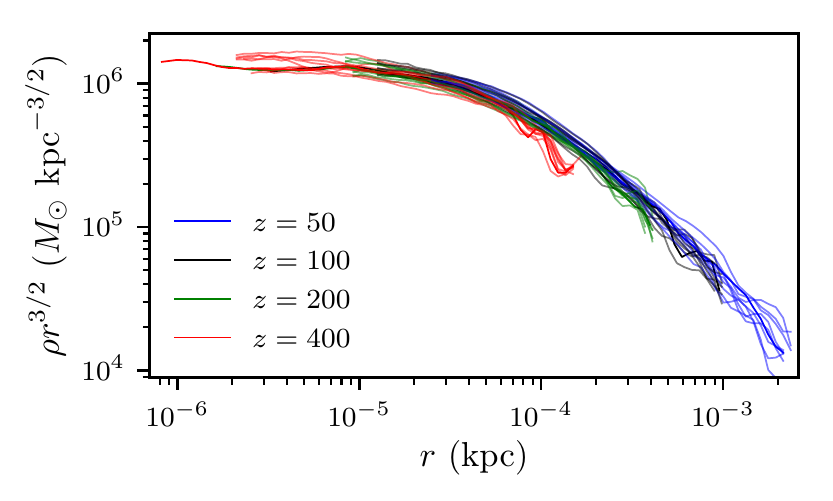}
	\caption{\label{fig:ucmhs} Radial density profiles at $z=400$, $z=200$, $z=100$, and $z=50$ of all nine UCMHs.  These density profiles are cut off above the virial radius at each redshift.  Evidently, all of our UCMHs possess similar density profiles to the one depicted in Fig.~\ref{fig:ucmh} (which is also plotted here).}
\end{figure}

We now study the spherically averaged density profiles of the UCMHs.  We simulated the UCMH in the primary simulation box at the highest particle density (see Appendix \ref{sec:conv} for details), so we first focus our study on that halo.  This halo has mass $M=31M_\odot$ at $z=100$, and Fig.~\ref{fig:ucmh} shows its density profile at $z=50$, $z=100$, $z=200$, and $z=400$ plotted in physical (not comoving) coordinates.  We first note that this halo clearly does not follow a $\rho\propto r^{-9/4}$ or similar single-power-law form, contradicting the assumption made in prior UCMH treatments.  We have conducted extensive convergence testing to confirm the validity of this result, as described in Appendix~\ref{sec:conv}.  The actual density profile is shallower, which will substantially reduce the observational signals of these halos, as we discuss in Sec. \ref{sec:constraints}.  However, the inner profile is still steeper than the $\rho\propto r^{-1}$ behavior of the NFW profile given by Eq.~(\ref{NFW}).  In fact, the inner density profile approaches $\rho\propto r^{-3/2}$, and the full density profile is fit well by the double-power-law form
\begin{equation}\label{Moore}
\rho(r) = \frac{\rho_s}{(r/r_s)^{3/2}(1+r/r_s)^{3/2}}
\end{equation}
which scales as $\rho\propto r^{-3/2}$ at small $r$ and $\rho\propto r^{-3}$ at large $r$.  We will call Eq.~(\ref{Moore}) the Moore profile due to its similarity to the form in Ref.~\cite{moore1999cold}.

Inner profiles $\rho\propto r^{-\gamma}$ with index $\gamma$ ranging from 1.3 to 1.5 have previously been observed in the smallest halos forming above a cutoff in the power spectrum
\cite{angulo2017earth,anderhalden2013density,ishiyama2010gamma,ishiyama2014hierarchical,polisensky2015fingerprints}, and Ref.~\cite{ogiya2017sets} found that the emergence of $\rho\propto r^{-3/2}$ is connected to the presence of a uniform-density core in the precursor density peak.  In this light, it is not surprising that $\rho\propto r^{-3/2}$ arises in our spiked power spectrum, since like a cutoff power spectrum, it lacks power below the scale of the spike and produces cored peaks in the primordial density field.  The physical origin of the $\rho\propto r^{-3/2}$ profile is not well understood, but it is known to be markedly less rotationally supported than the NFW profile \cite{ogiya2017sets}.

We next remark that the inner density profile does not appear to change in time: observe the remarkable concordance between the inner density profiles at different redshifts.  This behavior was noticed in the radial infall solution \cite{fillmore1984self,bertschinger1985self}, and it is explained in that context by the steepness of the potential well: newly accreted matter passes through the central regions too quickly to significantly affect the density there.  This effect is only enhanced in a three-dimensional picture, where newly accreted matter is likely to possess too much angular momentum to pass through the central parts of the halo at all.  The stability of the density profile in time is important to us for two reasons.  First, it allows us to use the measurement of the inner density profile at an early redshift as a proxy for the inner density profile at a later redshift, when the expansion of the comoving coordinates has brought the inner profile beyond our resolution limits (due to force softening; see Appendix \ref{sec:conv}).  In other words, we view the innermost points in Fig.~\ref{fig:ucmh}, present only at high $z$, as also representing the density profile at later $z$, e.g. $z=50$.  This argument allows us to claim that we have probed radii down to $10^{-3.5}r_\mathrm{vir}$ at $z=50$, where $r_\mathrm{vir}$ is the UCMH virial radius.  (If one does not accept this argument, we have still probed radii down to $10^{-2.5}r_\mathrm{vir}$ at $z=100$.)  Second, this stability means we can study the density profile at redshifts of order $z\sim 100$ and assume that---in the absence of disruptive events---the profile is the same today.  Observational signals can therefore be calculated using this profile (see Sec.~\ref{sec:constraints}).

Finally, we remark on the fitting parameters $\rho_s$ and $r_s$ of the Moore profile [Eq.~(\ref{Moore})] for the UCMH shown in Fig.~\ref{fig:ucmh}.  The scale radius $r_s$ that separates the $\rho\propto r^{-3/2}$ behavior from the $\rho\propto r^{-3}$ behavior appears to be set by the physical scale associated with the spike in the power spectrum at $z=1000$, obeying $r_s \simeq 0.7[(1+z)k_s]^{-1}$.  Similarly, the scale density $\rho_s$ is close to the background physical density at $z=1000$, obeying $\rho_s \simeq 30(1+z)^3\bar\rho_0$, where $\bar\rho_0$ is the background matter density today.  These correlations suggest that the $\rho\propto r^{-3/2}$ inner profile is set during the earliest stages of the halo's growth while the $\rho\propto r^{-3}$ outer profile grows during late accretion.  We will develop these ideas in more detail in Sec.~\ref{sec:later}.

All of these results come from the UCMH in the primary simulation run.  We also simulated eight other UCMHs, and we show the density profiles of all nine of them in Fig.~\ref{fig:ucmhs}.  All of these halos collapsed near $z=1000$, and there is clearly little deviation in the structure of these halos.  In particular, all of them exhibit the same $\rho\propto r^{-3/2}$ inner density profile, providing further evidence that the $\rho\propto r^{-9/4}$ pure power law density profile does not arise in a realistic formation scenario.

\subsection{Mass accretion}

\begin{figure}[t]
	\centering
	\includegraphics[width=\columnwidth]{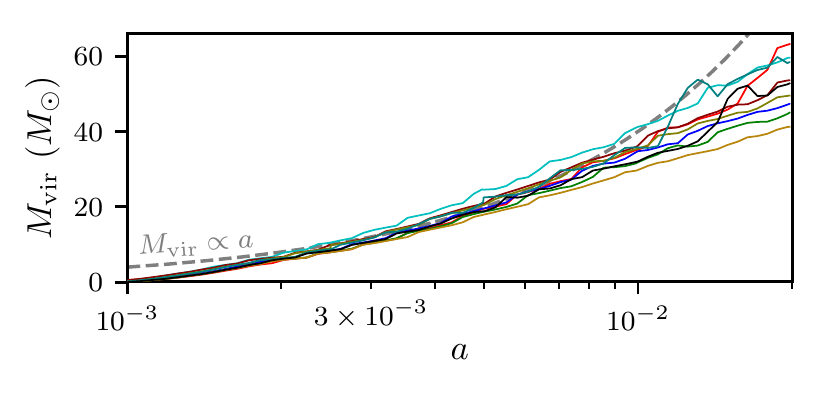}
	\caption{\label{fig:mass} The virial mass $M_\mathrm{vir}$ plotted against the scale factor $a$ for all nine UCMHs.  For $a\gtrsim \num{2.5e-3}$ ($z\lesssim 400$), the growth appears logarithmic, presenting as a straight line on this plot.  A $M_\mathrm{vir}\propto a$ reference curve is shown as a dashed line: these halos are growing more slowly than ${M_\mathrm{vir}\propto a}$.}
\end{figure}

We briefly remark on the mass accretion history of the UCMHs.  UCMHs have been previously assumed to grow as $M \propto a$ \cite{ricotti2009new}, but this is a result from radial infall theory \cite{bertschinger1985self}.  This theory describes an overdense region in an unperturbed background, which is very different from the Gaussian random field from which a realistic halo would form.

Figure~\ref{fig:mass} shows the growth of our UCMHs in virial mass $M_\mathrm{vir}$.  For $z\lesssim 400$ (${a\gtrsim \num{2.5e-3}}$), the mass of these halos appears to be logarithmic in $a$.  We do not claim that this logarithmic behavior necessarily continues to later times: halos that form from flatter power spectra have been observed to grow with redshift $z$ as $M\propto e^{-\alpha z}$ for some $\alpha$ \cite{wechsler2002concentrations}, which becomes slower than logarithmic, and halos in a spiked power spectrum could exhibit similar growth.  However, we have confirmed that these halos grow much more slowly than prior UCMH treatments have assumed.

\subsection{Mergers}\label{sec:merge}

\begin{figure}[t]
	\centering
	\includegraphics[width=\columnwidth]{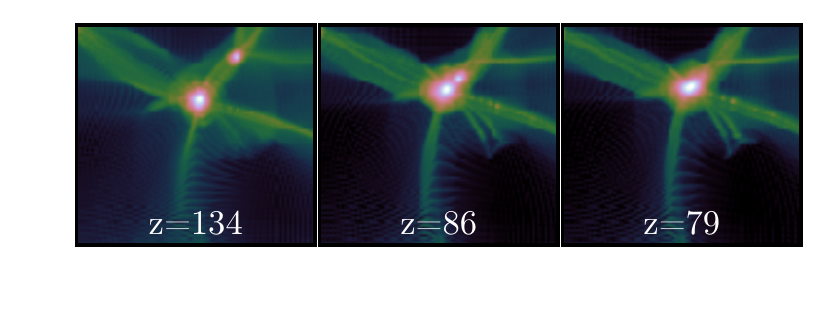}\\
	\vspace{-.6cm}
	\includegraphics[width=\columnwidth]{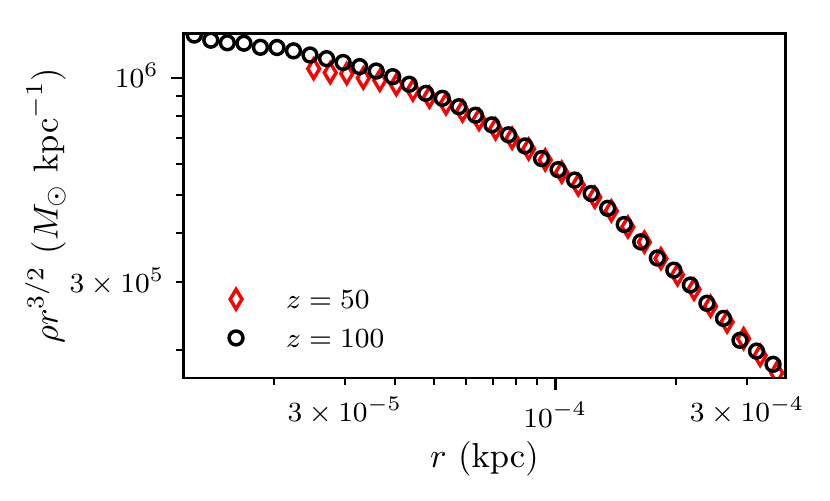}
	\caption{\label{fig:merge} A merger event and its result.  Top: A $(\SI{1.5}{kpc})^2\times$ 0.7 kpc (projected) region showing the first merger event experienced by the larger halo, which formed at $z\simeq 1000$.  Bottom: The change in the density profile of this UCMH as a result of the merger.  Mass is dispersed from the inner region.}
\end{figure}

We noted earlier that halo density profiles are expected to remain stable over time in the absence of disrupting events.  Halo mergers, however, are disruptive events and may be expected to alter the inner density profile.  In fact, this topic has been already explored in the context of the steeper inner profiles ($\rho\propto r^{-\gamma}$ with $\gamma>1$) that arise in the smallest halos above a cutoff in the power spectrum \cite{ogiya2016dynamical}.  Consecutive mergers cause these halos to relax toward shallower $\rho\propto r^{-1}$ inner density profiles.  However, these simulations used halos with concentration parameter $c=r_\mathrm{vir}/r_s \simeq 2$, where $r_\mathrm{vir}$ is the halo virial radius and $r_s$ is the scale radius.  Halos forming in a spiked power spectrum are sufficiently isolated that they may be expected to have concentration parameters of order 10 or higher by the time a merger takes place.  A systematic study of the effect of halo mergers on highly concentrated halos is beyond the scope of this paper, but we will briefly discuss in this section the effect of a merger on one of our UCMHs.

Three of our nine UCMHs underwent mergers between $z=100$ and $z=50$, with another two impending.  One such event occurring at $z\simeq 86$ is depicted in the upper panel of Fig.~\ref{fig:merge}.  The UCMH had concentration $c=12$ at this time.  The lower panel shows the density profile of this halo at $z=100$ and $z=50$ before and after the merger takes place, and we see that this event has been energetic enough to disperse mass out of the center of the halo and make the inner profile shallower.  Unfortunately, we do not have the resolution at these redshifts to determine the slope of the inner profile after the merger, but the fact that the density profile at $r<r_s$ is altered indicates that the stability we observed in Sec.~\ref{sec:density_profiles} does not hold after mergers.

\subsection{Other minihalos}\label{sec:later}

\begin{figure}[t]
	\centering
	\includegraphics[width=\columnwidth]{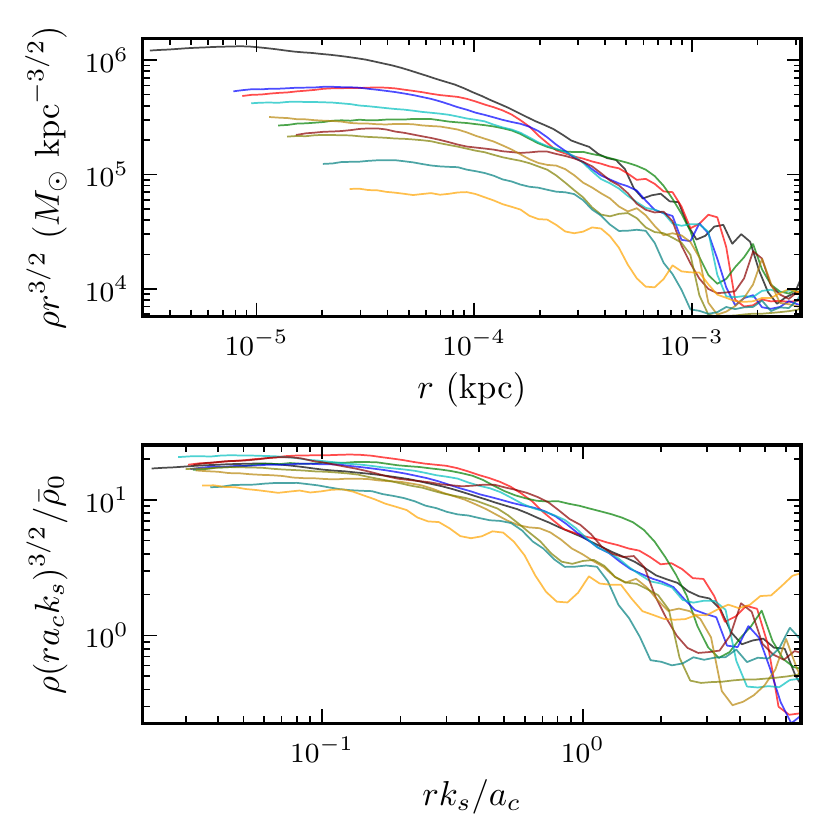}
	\caption{\label{fig:others} The density profiles of a sample of 10 halos in the same simulation box at $z=100$.  Top: The density profiles in physical coordinates.  Bottom: The density profiles scaled to each halo's formation time using Eqs. (\ref{rs_scale}) and~(\ref{rhos_scale}).}
\end{figure}

So far, we have studied only the exceptionally rare halos that form at $z\simeq 1000$.  In this section, we explore a sample of other halos in the simulation box shown in Fig.~\ref{fig:halo}.  We pick 10 halos, including the UCMH, with masses evenly distributed between 3 $M_\odot$ and 32 $M_\odot$ at $z=100$.  Figure~\ref{fig:others} shows the density profiles of these halos.  As we discussed in Sec.~\ref{sec:density_profiles}, we expect that each halo will obey
\begin{align}\label{rs_scale}
r_s &\propto a_\mathrm{c}k_s^{-1}\\
\label{rhos_scale}
\rho_s &\propto a_\mathrm{c}^{-3}\bar\rho_0,
\end{align}
where $a_\mathrm{c}$ is the scale factor at the halo's formation.  To test this hypothesis, we must determine $a_\mathrm{c}$ for each halo.  We do so using linear theory in the following way.  We find the earliest time at which \textsc{Rockstar} identifies the halo and map the location of the halo at this time onto the initial density grid.  Then we walk from this grid-cell to a local maximum in the density field by successively moving to the densest neighboring cell.  This local maximum is taken to be the amplitude $\delta_\mathrm{pk}$ of the protohalo peak.  Finally, we evolve the grid using linear theory, Eq.~(\ref{evo}), and find the time at which $\delta_\mathrm{pk} = 1.686$, the linear threshold for collapse.  The scale factor at this time is taken to be $a_\mathrm{c}$.

With a formation time $a_\mathrm{c}$ now associated with each halo, we test Eqs. (\ref{rs_scale}) and~(\ref{rhos_scale}) by plotting in Fig.~\ref{fig:others} the same density profiles with $\rho$ scaled to $a_c^{-3}\bar\rho_0$ and $r$ scaled to $a_c k_s^{-1}$ for each halo.  We find that the scatter in the density profiles is greatly reduced, with the bulk of the halos obeying
\begin{equation}\label{scaling1}
\rho_s r_s^{3/2} \simeq 17 \bar\rho_0 k_s^{-3/2} a_\mathrm{c}^{-3/2}
\end{equation}
($\rho_s r_s^{3/2}$ is the $r\ll r_s$ asymptote of $\rho r^{3/2}$ for a Moore profile).  The two halos lying farthest below this line have formed only slightly before the time $z=100$ at which we are seeing them, so it is plausible that their inner profiles are still growing.

We do not attempt to study the scaling of $\rho_s$ and $r_s$ separately because this requires fitting functional forms to the density profiles, which is unreliable with the resolution to which we are limited here.  However, $\rho_s r_s^{3/2}$ alone is a useful combination because it determines most of the annihilation signal of the halo (see Sec.~\ref{sec:constraints}).  Our ultimate goal is to predict halo density profiles from the power spectrum in order to place constraints thereon, and we find the spread in $\rho_s r_s^{3/2}$ to be well within a factor of 2 of Eq.~(\ref{scaling1}), which is promising.  However, our halo sample is small and we are biased by resolution toward larger halos.  We are also limited to a single power spectrum.  We will carry out in Paper~III a more systematic study of the density profiles of halos forming from spiked power spectra.

\subsection{Power spectrum with step}\label{sec:step}

\begin{figure}[t]
	\centering
	\includegraphics[width=\columnwidth]{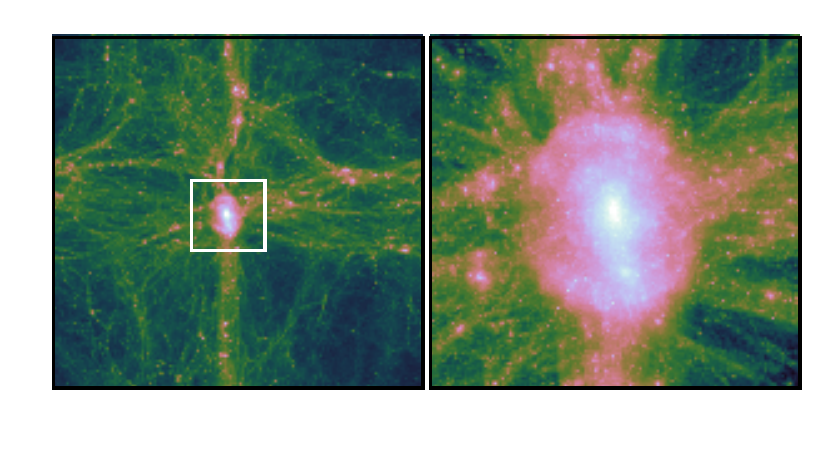}\\
	\vspace{-.4cm}
	\includegraphics[width=\columnwidth]{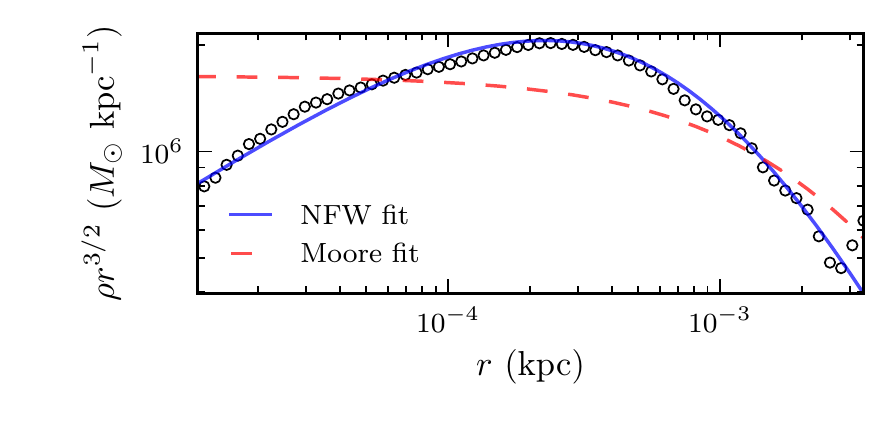}
	\caption{\label{fig:step} A halo at $z=100$ that collapsed at $z\simeq 1000$ from the step power spectrum shown in Fig.~\ref{fig:prim}.  Top: The projected full (7.4 kpc)$^3$ density field and an expanded picture of the 1.5 kpc cube surrounding the UCMH.  Bottom: The density profile of this halo.  It is fit well by the NFW profile.}
\end{figure}

We finally step away from the spiked power spectrum to verify that a picture with power evenly distributed across scales still produces NFW halos even in the UCMH scenario involving the early collapse of rare extreme overdensities.  We used the step power spectrum shown in Fig.~\ref{fig:prim} and prepared a set of initial conditions in a (7.4 kpc)$^3$ periodic box using the procedure described in Sec. \ref{sec:setup}.  Boxes were repeatedly generated until the $z=1000$ collapse criterion was met, which occurred after about $2300$ boxes.  We began the simulation run at $z=8\times 10^6$ and ended it at $z=100$; the resulting UCMH at $z=100$ is shown in Fig.~\ref{fig:step}.  It is evident from the density field that this is a very different picture from what we have seen with our spiked power spectrum.  The large-scale power has caused much of the mass within the box to collapse into the UCMH, while at the same time, the small-scale power has given this halo abundant substructure.

Figure~\ref{fig:step} also shows the radial density profile of this halo.  It follows the NFW form well, and does not fit the Moore form at all.  Moreover, we resolve an inner density profile that is at least as shallow as $\rho \propto r^{-1}$.  Even halos that collapse near $z\simeq 1000$ still possess the shallow inner profiles characteristic of hierarchical clustering.

A natural question to ask is how the density profiles behave in the transition between a spiked and a scale-invariant power spectrum.  A careful treatment is beyond the scope of this paper, but we will see in Paper~III that the answer is ultimately related to mergers.  As we discussed in Sec.~\ref{sec:merge}, mergers induce shallowing of the inner density profile toward $\rho\propto r^{-1}$.  Meanwhile, mergers occur more frequently when the power spectrum spike is wider, culminating in the hierarchical clustering characteristic of conventional power spectra.  These concepts explain, at least qualitatively, the shift from $\rho\propto r^{-3/2}$ to $\rho\propto r^{-1}$ inner profiles when the spike in the power spectrum is replaced by a step.

As a final remark, we have found between the spiked and step power spectra that UCMHs develop the same density profiles as halos that form at much later times.  The spike produces UCMHs with similar density profiles to those of the smallest halos forming above a free-streaming cutoff, while the step produces UCMHs with density profiles resembling those of the galactic halos created by hierarchical clustering.  In retrospect, this is not surprising.  Ref.~\cite{ricotti2009new} conceived of UCMHs as the late stage of rare non-Gaussian density fluctuations, so they assumed a conventional (unenhanced) power spectrum when calculating the velocity dispersion at $z\simeq 1000$.  The small velocity dispersion that resulted was the basis for the argument that radial infall theory would apply, but this velocity dispersion would be increased by any power spectrum enhancement.  There is no difference, aside from the emerging dominance of a radiation or dark energy component, between halos forming from a boosted power spectrum at early times and halos forming from a conventional power spectrum at late times.  However, the velocity dispersion is not the only obstacle to the $\rho\propto r^{-9/4}$ profile.  As we noted in Paper~I, this profile results specifically from the collapse of an overdense region in an unperturbed background, which is not an instance of a peak that forms in a Gaussian random field.  We will revisit this topic in Paper~III when we explore the relationship between a collapsed halo and its precursor density peak.

\section{Constraining the power spectrum}\label{sec:constraints}

UCMHs have been employed to constrain the primordial power spectrum through nonobservation of their predicted signals in a variety of contexts.  For thermal-relic dark matter models, such as the weakly interacting massive particle (WIMP) model \cite{jungman1996supersymmetric,bergstrom2000non,bertone2005particle}, the dark matter annihilation rate is greatly increased by the compactness of the assumed $\rho\propto r^{-9/4}$ density profile.  The strongest constraints therefore come from nonobservation of the strong gamma-ray \cite{josan2010gamma,bringmann2012improved,nakama2018constraints} or neutrino \cite{yang2013neutrino,yang2017tau} signals that are expected from WIMP annihilation within such dense clumps.  These annihilation signals would also lead to other observable effects, such as heating of the intergalactic medium \cite{zhang2011impact,yang2016contributions} and galactic gas \cite{clark2017heating} and interactions with the CMB or other background photons \cite{yang2011new,yang2013contribution,beck2018through}.
The primordial power spectrum can also be constrained by searching for UCMHs using astrometric microlensing \cite{li2012new} or macrolens distortions \cite{zackrisson2013hunting} or by constraining the UCMH abundance using pulsar timing arrays \cite{clark2015investigatingI,*clark2016erratumI,clark2015investigatingII,*clark2017erratumII}.

However, with the exception of Ref.~\cite{nakama2018constraints}, all of these works used only minihalos that form at $z\gtrsim 1000$ and assumed these halos possessed the $\rho\propto r^{-9/4}$ density profile.  We showed in Sec.~\ref{sec:ucmh} that minihalos forming in an enhanced power spectrum, even UCMHs forming at $z\gtrsim 1000$, develop significantly shallower profiles.  We also found that younger minihalos possess the same density profiles as the oldest ones.  In this section, we explore the impact of this discovery.  The observational signatures of UCMHs forming at $z\gtrsim 1000$ are weakened by the shallower density profile, but our analysis is now able to include minihalos forming at $z<1000$.  As we saw in Fig.~\ref{fig:halo}, these younger minihalos are far more abundant than the rare UCMHs.

Broadening to the entire population of minihalos brings new challenges.  Minihalo-minihalo mergers will reduce the minihalo count and alter their density profiles \cite{ogiya2016dynamical}, and tidal interactions within galactic structures will have more impact on the shallower density profiles \cite{berezinsky2008remnants}.  These considerations are beyond the scope of this paper, but to motivate further study, we calculate in this section how the new minihalo picture directly alters previous constraints on the power spectrum derived from UCMHs.  To this end, we focus on the upper bound derived by Bringmann, Scott, and Akrami \cite{bringmann2012improved} (hereafter BSA) based on the gamma-ray signal from WIMP annihilation within UCMHs.  In our calculation, we adopt our new minihalo model from Sec.~\ref{sec:later} but otherwise replicate BSA's calculation as closely as possible.  In particular, we employ the same Fermi-LAT data and, like BSA, when deriving bounds from diffuse emission, we consider only a Galactic contribution and neglect the possibility of improving constraints by including extragalactic sources (e.g., Refs.~\cite{yang2011constraints,nakama2018constraints}).

The derivation of a constraint on the power spectrum using WIMP annihilation in minihalos proceeds in three steps:
\begin{enumerate}
  \item The annihilation signal of a minihalo is calculated.
  \item A constraint on the number density of minihalos is calculated from the nonobservation of such an annihilation signal.
  \item The number density constraint is converted into a constraint on the primordial power spectrum using the statistics of Gaussian random fields.
\end{enumerate}
In past studies, such as BSA, UCMHs are assumed to collapse at $z\simeq 1000$, so the UCMH luminosity is solely a function of its size.  We now have the machinery to study minihalos collapsing at any redshift, so we calculate the minihalo luminosity $L$ as a function of both its formation time and the scale of the density fluctuation that sourced it.

\subsection{Halo luminosity}

We assume that the density profile of a minihalo follows the Moore fitting form given by Eq.~(\ref{Moore}).  In addition, we found in Sec.~\ref{sec:ucmh} that the Moore fitting parameters $r_s$ and $\rho_s$ can be predicted from the halo formation time as
\begin{align}
r_s = f_1 k_s^{-1} a_c
\nonumber\\
\rho_s = f_2\bar\rho_0 a_c^{-3},
\end{align}
where $\bar\rho_0 = \Omega_c\rho_\mathrm{crit}$ is the background dark matter density today, $k_s$ is the (comoving) wave number associated with the spike in the power spectrum, and $a_c$ is the halo formation time in a spherical collapse model.  The coefficients $f_1$ and $f_2$ are determined from simulations; they possess some scatter between halos, but we will neglect that scatter for the purpose of this calculation.  The annihilation signal of the Moore profile depends dominantly on the combination $f_2^2 f_1^3$ and only logarithmically on $f_2$ alone, so we will use $f_2 f_1^{3/2} = 17$ from Eq.~(\ref{scaling1}), which was derived from a sample of halos forming at different times, along with the more approximate value $f_2\simeq 30$ derived from the UCMHs alone.

The gamma-ray signal $L$ of a halo with density profile $\rho(r)$ may be calculated as
\begin{equation}\label{lum}
L = 4\pi g\int_0^{R}r^2\rho^2(r)\mathrm{d}r,
\end{equation}
where $R$ is the radius of the halo and $g$ is a factor related to the annihilation mechanism.  For threshold photon energy $E_\mathrm{th}$,
\begin{equation}\label{g}
g = \sum_k\int_{E_\mathrm{th}}^{m_\chi} E\frac{\mathrm{d}N_k}{\mathrm{d}E} \mathrm{d}E
\frac{\langle\sigma_k v\rangle}{2m_\chi^2},
\end{equation}
where $\mathrm{d}N_k/\mathrm{d}E$ is the differential photon yield of the $k$th annihilation channel and $\langle \sigma_k v\rangle$ is its cross section.  Equation~(\ref{g}) describes the energy flux; for the photon flux, the factor of $E$ is removed from the integrand.

For now, we keep our calculations model-independent and return to Eq.~(\ref{lum}).  Equation~(\ref{lum}) diverges for a $\rho\propto r^{-3/2}$ profile, but this implies that annihilations would have smoothed out the central cusp within some small radius.  We use the standard estimate \cite{berezinsky1992distribution}
\begin{equation}
\rho_\mathrm{max} = \frac{m_\chi}{\langle \sigma v \rangle (t-t_\mathrm i)}
\end{equation}
for the maximum density at time $t$ in a structure that formed at $t_\mathrm i$, where $m_\chi$ is the mass of the WIMP and $\langle\sigma v\rangle$ is its thermally averaged velocity-weighted cross section (in the zero-velocity limit).  Note that $t-t_\mathrm i \simeq t$ today if halo formation occurs at $z\gtrsim 10$, and we will see in Sec.~\ref{sec:constraint_discussion} that this is true for the minihalos relevant to power spectrum constraints.  Thus, we take $t-t_\mathrm i$ to be the age of the universe today, making $\rho_\mathrm{max}$ the same for all minihalos.  For a canonical WIMP with $\langle\sigma v\rangle = \SI{3e-26}{cm^3 s^{-1}}$ and $m_\chi=\SI{1}{TeV}$, $\rho_\mathrm{max}\sim 10^{16}\bar\rho_0$.

We now evaluate Eq.~(\ref{lum}) for a Moore profile as given by Eq.~(\ref{Moore}) modified to have maximum density $\rho_\mathrm{max}$.  The choice of radius $R$ has negligible impact as long as $R > r_s$, so taking $R \to \infty$, we obtain
\begin{equation}\label{Moorelum1}
L = 4\pi g\rho_s^2 r_s^3\left[\frac{1}{3}+\ln(1+D)-\frac{3+2D^{-1}}{2(1+D^{-1})^2}\right],
\end{equation}
where $D \equiv (\rho_\mathrm{max}/\rho_s)^{2/3} = (f_2^{-1}\rho_\mathrm{max}/\bar\rho_0)^{2/3}a_c^{2}$.  For halos collapsing at $z\lesssim 1000$, we find that $D \gtrsim 10^4 \gg 1$ for a canonical WIMP, so Eq.~(\ref{Moorelum1}) simplifies to
\begin{equation}\label{Moorelumshort}
L \simeq B k_s^{-3} a_c^{-3} \ln (\beta a_c)
\end{equation}
with
\begin{equation}\label{Moorelumcoef}
B\equiv 8\pi g f_2^2f_1^3\bar\rho_0^2,
\ \ \ 
\beta\equiv e^{-7/12}\left(\frac{\rho_\mathrm{max}}{f_2\bar\rho_0}\right)^{1/3}.
\end{equation}
$B$ and $\beta$ are independent of the scale $k_s$ of the spike in the power spectrum, and since we are neglecting scatter in $f_1$ and $f_2$, they are the same for all halos.  We remark that since $\beta \sim 10^5$, the logarithmic dependence on $a_c$ is weak for $a_c\gtrsim 10^{-3}$.  If halos relax to a $\rho\propto r^{-1}$ inner profile due to mergers or other disruptive dynamics, a similar calculation with the NFW profile yields
$L \simeq B k_s^{-3} a_c^{-3}/6$, which is smaller by a factor of $\sim\! 30$ for formation time $z\sim 100$.  In this case, Eq.~(\ref{lum}) converges, so the effect of $\rho_\mathrm{max}$ is negligible.

\subsection{Halo abundance}

We use observational detection limits to constrain the minihalo number density based on the luminosity we computed above.  Following BSA, we employ two approaches that utilize different observations and yield different constraints.  First, we treat the minihalos as point sources and use their nonobservation to constrain their number density.  Next, we consider the diffuse background flux from minihalos within the Milky Way and use the observed background gamma-ray flux to constrain the minihalo number density.

\subsubsection{Point sources}

The gamma-ray flux $\mathcal{F}$ from a point source is related to its luminosity $L$ and distance $d$ by $\mathcal{F} = L/(4\pi d^2)$.  If our detecting instrument has flux sensitivity $\mathcal{F}_\mathrm{min}$ to point sources, then this imposes a maximum observable distance $d_\mathrm{obs} = \sqrt{L/(4\pi \mathcal{F}_\mathrm{min})}$ corresponding to the observable volume
\begin{equation}\label{Vobs}
V_\mathrm{obs} = \frac{1}{3\sqrt{4\pi}}\frac{L^{3/2}}{F_\mathrm{min}^{3/2}}.
\end{equation}
If $V_\mathrm{obs}$ were the same for all halos, then the expected number of observable objects would be $\lambda = n V_\mathrm{obs}$, where $n$ is the halo number density, and we could use Poisson statistics to constrain $n$ from our knowledge of $V_\mathrm{obs}$.  However, in our model, $L$, and hence $V_\mathrm{obs}$, is a function of the formation time $a_c$ of the minihalo.  Instead of the total number density $n$, we must consider the differential number density $\mathrm{d}n/\mathrm{d}a_c$ of minihalos forming at $a=a_c$.  The expected number of observable minihalos is now
\begin{equation}\label{ptexp}
\lambda = \int_0^1 \mathrm{d} a_c \left(\frac{\mathrm{d}n}{\mathrm{d}a_c}\right)_\mathrm{obs} V_\mathrm{obs}(a_c),
\end{equation}
where we write $(\mathrm{d}n/\mathrm{d}a_c)_\mathrm{obs}$ to clarify that we are referring to the number density of minihalos within $V_\mathrm{obs}$, which in general differs from the cosmological mean $\mathrm{d}n/\mathrm{d}a_c$.

From Poisson statistics, the probability that there is at least one observable object is $P(N_\mathrm{obs}>0) = 1-e^{-\lambda}$.  If the confidence level associated with the flux threshold $\mathcal{F}_\mathrm{min}$ is $x$, then the probability of observing at least one object is $P_\mathrm{obs} = x(1-e^{-\lambda})$.  If we observe no objects, an upper bound on $\lambda$ with confidence level $y$ is obtained by setting $P_\mathrm{obs} \leq y$, implying $\lambda \leq -\ln(1-y/x)$.  Combining this result with Eqs. (\ref{Vobs}) and~(\ref{ptexp}), we find
\begin{equation}\label{n_pt}
\int_0^1 \mathrm{d} a_c \left(\frac{\mathrm{d}n}{\mathrm{d}a_c}\right)_\mathrm{obs} L^{3/2}(a_c) \leq -3\sqrt{4\pi}\ln(1-y/x) \mathcal F_\mathrm{min}^{3/2},
\end{equation}
which gives us the prescription for constraining the local number density of minihalos based on the nonobservation of point sources.  Due to the dependence of a minihalo's luminosity on its formation time, we constrain a formation time-weighted density instead of a total UCMH density.

To complete the calculation, we need to relate $(\mathrm{d}n/\mathrm{d}a_c)_\mathrm{obs}$ to the cosmological mean $\mathrm{d}n/\mathrm{d}a_c$ that is predicted by the power spectrum.  To do this, we assume that the spatial distribution of minihalos is proportional to that of dark matter at large\footnote{Galactic tides and other disruptive processes would realistically alter the spatial distribution of minihalos, but we neglect them here.}; that is, $n(\vec x)\propto \rho(\vec x)$.  We define $\mu(d) \equiv 3M(d)/(4\pi d^3\bar\rho_0)$ as the ratio of the dark matter mass $M(d)$ contained within distance $d$ from Earth to the cosmological mean dark matter mass contained within an equal volume.  Then the mean minihalo number density within $d_\mathrm{obs}$ is related to the cosmological mean by the factor $\mu(d_\mathrm{obs})$, implying
\begin{equation}\label{dnda_obs}
\left(\frac{\mathrm{d}n}{\mathrm{d}a_c}\right)_\mathrm{obs}
=
\mu\!\left(\sqrt\frac{L(a_c)}{4\pi \mathcal{F}_\mathrm{min}}\right)
\frac{\mathrm{d}n}{\mathrm{d}a_c}.
\end{equation}
We evaluate $\mu(d)$ in Appendix~\ref{sec:ptunif} and plot it in Fig.~\ref{fig:mu} assuming an NFW profile for the Milky Way with parameters from Ref.~\cite{battaglia2005radial}.  For $d_\mathrm{obs} \lesssim \SI{8}{kpc}$, the distance to the Galactic center, $\mu(d_\mathrm{obs})\simeq 2\times 10^5$ is approximately constant.

\begin{figure}[t]
	\centering
	\includegraphics[width=\columnwidth]{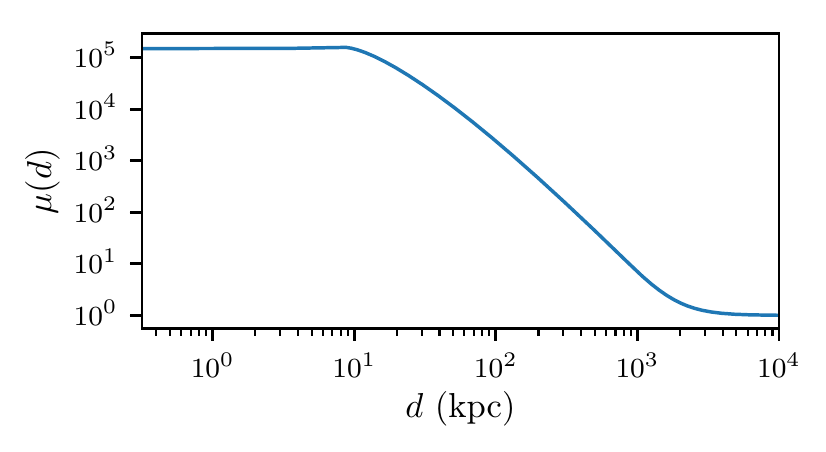}
	\caption{\label{fig:mu} The ratio $\mu(d)=3M(d)/(4\pi d^3\bar\rho_0)$ of the dark matter mass $M(d)$ within distance $d$ of Earth to the cosmological mean dark matter mass in an equal volume.  An NFW profile is assumed for the Milky Way with parameters from Ref.~\cite{battaglia2005radial}.  Minihalos are assumed to follow this spatial distribution.}
\end{figure}

\subsubsection{Diffuse flux}

The calculation is simpler for the case of a diffuse gamma-ray flux.  If $\mathrm d \mathcal{F}/\mathrm d \Omega$ is the upper bound on the observed differential gamma-ray flux that can be attributed to minihalos, then we can relate this to the differential flux summed over all minihalos along the line of sight,
\begin{equation}\label{n_diff0}
\frac{\mathrm d \mathcal F}{\mathrm d \Omega} \geq
\int_0^\infty s^2 \mathrm d s \int_0^1 \mathrm d a_c \frac{\mathrm d n}{\mathrm d a_c}\frac{\rho(s)}{\bar\rho_0} \frac{L(a_c)}{4\pi s^2},
\end{equation}
where $s$ is the line-of-sight distance.
Here we have inserted the factor $\rho(s)/\bar\rho_0$ to account for the Milky Way density field at distance $s$ from Earth.  Following BSA, we are only interested in the Galactic contribution to the diffuse flux, so we truncate the density field beyond the Milky Way, eliminating the need to redshift distant sources.  The minihalo abundance constraint from the diffuse flux at angle $\theta$ to the Galactic center now becomes
\begin{equation}\label{n_diff}
\int_0^1 \mathrm d a_c \frac{\mathrm d n}{\mathrm d a_c} L(a_c)
\leq
\frac{4\pi}{K(\theta)}\frac{\mathrm d \mathcal F}{\mathrm d \Omega},
\end{equation}
where 
\begin{equation}\label{n_diff_k}
K(\theta)\equiv\int_0^\infty \mathrm d s \frac{\rho_\mathrm{MW}(\sqrt{s^2+r_0^2-2 s r_0 \cos\theta})}{\bar\rho_0},
\end{equation}
$\rho_\mathrm{MW}(r)$ is the Milky Way density profile, and $r_0$ is the solar orbital radius.

\subsection{The power spectrum}\label{sec:powerspectrum}

Finally, we must find the relationship between $\mathrm{d}n/\mathrm{d}a_c$ and the power spectrum $\mathcal{P}(k)$.  The standard way to relate a halo population to a power spectrum is Press-Schechter theory \cite{press1974formation}.  For any power spectrum with a small-scale cutoff, including a spiked power spectrum, it is necessary to employ a sharp $k$-space smoothing filter to avoid overpredicting structure below the cutoff scale \cite{barkana2001constraints,benson2012dark,schneider2013halo}.  However, there are additional challenges in adapting Press-Schechter theory to our purposes.  Halo formation times can be obtained from the conditional mass function \cite{lacey1993merger}, but these yield the average formation time of progenitor halos.  It is necessary to construct merger trees to study the first progenitor.  Also, Press-Schechter theory always destroys a halo when two halos merge.  For unequal-mass mergers, a remnant of the smaller halo is generally expected to survive as a subhalo with its central structure intact \cite{berezinsky2008remnants}.

For our calculation, we employ a more direct approach.  Bardeen, Bond, Kaiser, and Szalay \cite{bardeen1986statistics}, hereafter BBKS, formulated a description of the statistics of peaks in a Gaussian random field.  In this approach, each peak in the primordial density field is to be identified with a halo at late times.  Spiked power spectra are very natural arenas for peak theory because they possess finite integrated power and generate peaks around a particular scale, so it is not necessary to use any smoothing filter.  To represent a spike centered at wave number $k_s$, we consider a delta-function matter power spectrum of the form ${\mathcal P (k) \propto D(a)^2 k_s \delta(k-k_s)}$, where $D(a)$ is the linear growth function.  We will see in Sec.~\ref{sec:constraint_discussion} that the minihalos contributing to our power spectrum constraint form in matter domination, so $D(a)=a$, and we may write
\begin{equation}\label{pkdelta}
\mathcal P (k) = \mathcal{A} a^2 k_s \delta(k-k_s),
\end{equation}
where $\mathcal{A}$ parametrizes the integrated area of the spike.  We use the BBKS formalism to calculate the number density of peaks with $\delta>\delta_c$, where ${\delta_c=1.686}$ is the linear collapse threshold.  The identification of these peaks with halos leads to a number density $n$ that increases in time solely due to halo formation, implying we can differentiate it with respect to scale factor $a$ to obtain the distribution of halos by formation time.  A disadvantage to this procedure is that minihalo-minihalo mergers are not automatically accounted for and must be handled separately, a task that is beyond the scope of this paper\footnote{However, as we will see in Paper~III, mergers become rare as the power spectrum spike is narrowed.  With a delta-function spike in the primordial power spectrum, we suspect that they are negligible.}.

As detailed in Appendix~\ref{sec:bbks}, we obtain
\begin{equation}\label{dnda}
\frac{\mathrm{d}n}{\mathrm{d}a_c}= \frac{k_s^3}{a_c}\  h\!\left(\frac{\delta_c}{\mathcal{A}^{1/2}a_c}\right),
\end{equation}
where $h(\nu)$ is the distribution of peak heights given by Eq.~(\ref{dnda_func}) (see Fig.~\ref{fig:dnda_h}).  With the minihalo signal given by Eq.~(\ref{Moorelumshort}), the abundance constraints given in Eqs. (\ref{n_pt}) and~(\ref{n_diff}), and this relation between $\mathrm{d}n/\mathrm{d}a$ and the power spectrum, we can place an upper bound on the amplitude $\mathcal{A}$ of the spike in the matter power spectrum.  The final step is to convert this bound into a bound on the primordial curvature power spectrum.  We adopt a similar delta-functional form for the primordial power spectrum, 
\begin{equation}
\mathcal{P}_\zeta(k)=\mathcal{A}_0 k_s \delta(k-k_s),
\end{equation}
with amplitude $\mathcal{A}_0$.  The transfer function given by Eq.~(\ref{deltaMD}) converts the bound on $\mathcal{A}$ into a bound on $\mathcal{A}_0$.

To carry out the calculation, we assume a canonical WIMP with cross section $\langle\sigma v\rangle = \SI{3e-26}{cm^3 s^{-1}}$ and mass $m_\chi = \SI{1}{TeV}$ that annihilates into $b\bar b$ pairs.  We take the Fermi-LAT point-source sensitivity for energies above 100 MeV to be $\mathcal{F}_\mathrm{max}=\SI{4e-9}{cm^{-2} s^{-1}}$ for a $5\sigma$ detection, and we set $y=0.95$ in Eq.~(\ref{n_pt}) for a 95\% confidence limit.  For the diffuse flux, we use $\mathrm{d}\mathcal{F}/\mathrm{d}\Omega=\SI{1.2e-5}{GeV cm^{-2} s^{-1} sr^{-1}}$ as the $2\sigma$ limit (with systematic error alone) in the energy flux from the Galactic poles as measured by Fermi-LAT \cite{nolan2012fermi}.  Finally, we take the Milky Way to have an NFW density profile with parameters determined in Ref.~\cite{battaglia2005radial}.  All of these choices are picked solely for parity with BSA, and further detail can be found there.

\begin{figure}[t]
	\centering
	\includegraphics[width=\columnwidth]{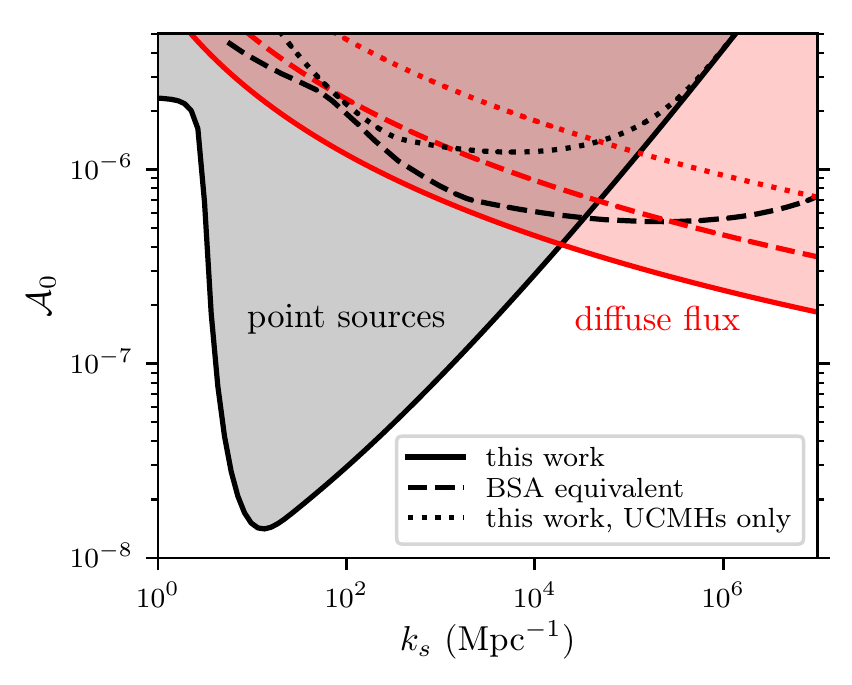}
	\caption{\label{fig:constraint} The upper bound on the integrated area $\mathcal{A}_0$ of a spike in the primordial curvature power spectrum centered at scale wave number $k_s$.  Black curves use point sources, while red curves employ the diffuse flux.  The shaded regions are ruled out in the new minihalo picture with shallower density profiles.  The dashed lines show the corresponding constraints in the old UCMH model calculated using the abundance constraints in BSA.  As another comparison, the dotted lines show the constraints using shallower density profiles while still restricting to UCMHs forming at $z\geq 1000$.  While the new density profiles slightly weaken the upper bound, the inclusion of all minihalos ends up leading to stronger constraints.}
\end{figure}

Figure~\ref{fig:constraint} shows the resulting upper bound on the integrated area $\mathcal A_0$ of a spike in the primordial curvature power spectrum if the spike is located at wave number $k_s$.  We show the constraints from point sources and diffuse flux separately, and the shaded regions are forbidden.  We wish to compare this constraint to the upper bound derived in BSA under the UCMH picture, but BSA assumed a locally scale-invariant power spectrum for their analysis.  Therefore, we employ the UCMH abundance constraints in BSA to derive a constraint on the spiked power spectrum of Eq.~(\ref{pkdelta}).  This calculation is detailed in Appendix~\ref{sec:bsa}, and the results are plotted on Fig.~\ref{fig:constraint} as dashed lines.  Evidently, new minihalo constraints can be stronger than old UCMH constraints despite employing shallower density profiles.  For comparison, we also show as dotted lines the upper bounds that employ the shallower density profiles while restricting to UCMHs forming by $z\geq 1000$.  These bounds are calculated by altering the upper limit of the integrals in Eqs. (\ref{n_pt}) and~(\ref{n_diff}).  We noted in Paper~I that the shallower density profiles reduce the signal from each halo by a factor of 200, and we see now that this reduction weakens the upper bound on the power spectrum by roughly a factor of two\footnote{In the next section, we discuss why power spectrum constraints derived from UCMHs are so insensitive to reductions in the UCMH signal.  This feature is a consequence of the restriction to halos forming at $z\geq 1000$ and is no longer applicable once all minihalos are included.}.  The inclusion of all minihalos, instead of only the rare UCMHs that form by $z=1000$, more than compensates for this loss.

\subsection{Discussion}\label{sec:constraint_discussion}

\begin{figure}[t]
	\centering
	\includegraphics[width=\columnwidth]{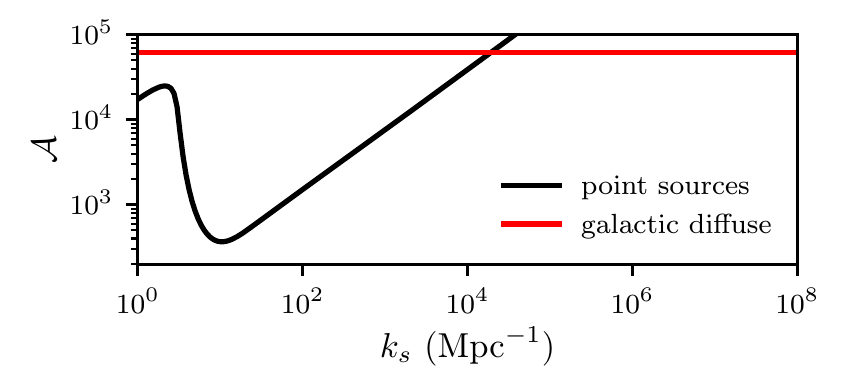}
	\caption{\label{fig:constraint0} The constraint on the integrated area $\mathcal{A}$ of the spiked matter power spectrum Eq.~(\ref{pkdelta}).  Note that this is not the primordial power spectrum; see Fig.~\ref{fig:constraint} for that constraint.}
\end{figure}

To develop a better understanding of the power spectrum constraints in this new minihalo picture, we specialize to the diffuse flux and to point sources in the small-object limit where $d_\mathrm{obs}$ is sufficiently short that ${\mu(d_\mathrm{obs})\equiv \mu}$ is constant\footnote{The small-object limit produces the power-law branch of the point-source constraint in Fig.~\ref{fig:constraint0}, implying that this limit corresponds to $k_s\gtrsim 20$ Mpc$^{-1}$.}.  In these cases, it is possible to derive the analytic constraints (see Appendix~\ref{sec:bbks})
\begin{align}\label{pt_muconst}
\mathcal{A}&\left(\ln\frac{\beta\delta_c}{\mathcal{A}^{1/2}}\right)^{2/9}\left[\left(\ln\frac{\beta\delta_c}{\mathcal{A}^{1/2}}\right)I_{3/2}-J_{3/2}\right]^{4/9}
\nonumber\\
&\leq
\left(\frac{-3\sqrt{4\pi}\ln(1-y/x)}{\mu}\right)^{4/9}\left(\frac{\delta_c^3 k_s \mathcal F_\mathrm{min}}{B}\right)^{2/3}
\end{align}
for point sources in a uniform field with $\mu$ times the background density and
\begin{align}\label{diffuse}
\mathcal{A}\left[\left(\ln\frac{\beta\delta_c}{\mathcal{A}^{1/2}}\right)I_1-J_1\right]^{2/3}
\leq
\left(\frac{4\pi\delta_c^3}{K(\theta)B}\frac{\mathrm d \mathcal F}{\mathrm d \Omega}\right)^{2/3}
\end{align}
for diffuse sources.  Here, $I_{3/2} = 0.228$, $J_{3/2} = 0.370$, $I_1 = 0.0477$, and $J_1 = 0.0478$ are different moments of the peak height distribution $h(\nu)$.

We first note that if we neglect logarithms\footnote{Using Fig.~\ref{fig:constraint0}, $\beta\sim 10^5 \gg\mathcal{A}^{1/2}$, so the logarithmic dependence of Eqs. (\ref{pt_muconst}) and~(\ref{diffuse}) on $\mathcal{A}$ is weak.}, the constraint on $\mathcal A$ is proportional to $B^{-2/3}$ and hence to the $-2/3$ power of the WIMP annihilation rate within minihalos [see Eq.~(\ref{Moorelumshort})].  This relationship implies that the upper bound on $\mathcal A$ is highly sensitive to the WIMP model.  For example, if the annihilation cross section $\langle\sigma v\rangle$ were increased by a factor of 8, the upper bound on $\mathcal A$ would be reduced by a factor of 4.  This behavior is a stark contrast to that of constraints in the old UCMH picture, which exhibit a very weak dependence on WIMP model (see BSA Fig.~5).

The same distinction arises when considering the observational flux constraint $\mathcal F_\mathrm{min}$ or $\mathrm d \mathcal F/\mathrm d \Omega$.  The upper bound on $\mathcal{A}$ is more sensitive to these observational constraints in the new minihalo picture than in the old UCMH picture.  Therefore, improved observational limits are far more valuable in the new picture.  This property is also responsible for how, as depicted in Fig.~\ref{fig:constraint}, the point-source constraint in the new picture exhibits markedly stronger $k_s$-dependence: larger objects are more visible, and this heightened visibility now significantly strengthens the upper bound on the power spectrum on the corresponding scales.  Likewise, we saw in the last section that reducing the UCMH gamma-ray signal by a factor of 200 only weakens the UCMH-derived power spectrum bounds by a factor of two.  A similar change to the luminosity of all minihalos would weaken the bounds in the new picture by a factor of 34.

These differences in sensitivity can be understood in the following way.  Upper bounds on minihalo abundance ($f$ in BSA; $n$ or $\mathrm{d}n/\mathrm{d}a_c$ here) are always highly sensitive to minihalo signals and observational flux constraints, whether we restrict to UCMHs or not; compare Eqs. (\ref{n_pt}) and~(\ref{n_diff}) to BSA Eqs. (26) and~(29).  However, the sensitivity of a power spectrum bound to these abundance constraints depends on the types of minihalos that contribute.  In the old UCMH picture, constraints were dominated by halos forming from initial overdensities that correspond to $5\sigma$-$6\sigma$ fluctuations.  These peaks are so far out in the Gaussian tail of the density distribution that altering their abundance only marginally changes the distribution's spread\footnote{The (differential) abundance of a density excess $\delta$ is proportional to $\exp\!\left(-\frac{1}{2}\delta^2/\sigma^2\right)$ if $\delta$ is distributed with spread $\sigma$.  If $\delta/\sigma$ is large, then a large change in the abundance---the quantity constrained by observations---corresponds to a small change in $\sigma$, which sets the power spectrum normalization.  (This is just an illustration: to be precise, we should use the cumulative distribution function.)}.  In the new minihalo picture, constraints are influenced by the bulk of the peaks, so an alteration to the abundance of these peaks now changes the spread of the distribution more drastically.  We also remark on another consequence of this difference in statistics: the constraints will no longer be as sensitive to possible small deviations from Gaussianity that would significantly affect the tails of the distribution \cite{shandera2013number}.

The influence of the peak population on the power spectrum constraint is encoded in the moments $I_{3/2}$, $J_{3/2}$, $I_1$, and $J_1$ of the peak distribution.  These are integrals over peak height $\nu=\delta/\sigma$, and their integrands exhibit most of their support between $\nu=2$ and $\nu=4$ (see Appendix~\ref{sec:bbks}).  Consequently, the integrals in Eqs. (\ref{n_pt}) and~(\ref{n_diff}) that determine the upper bounds on the power spectrum are dominated by peaks with amplitudes between $2\sigma$ and $4\sigma$, which confirms the difference in statistics from the old UCMH picture.  We can also use this information to find the formation times of the corresponding halos.  The upper bound on $\mathcal{A}$, which parametrizes the matter power spectrum, is shown in Fig.~\ref{fig:constraint0} and lies between \num{3e2} and \num{6e4}.  The root-mean-squared density variance of the spiked power spectrum is $a\mathcal{A}^{1/2}$ at scale factor $a$, implying that the collapse time $a_c$ of a peak with amplitude $\nu$ is $a_c=\delta_c/(\nu \mathcal{A}^{1/2})$.  It follows that peaks contributing significantly to the power spectrum constraint would have formed between $z=20$ and $z=600$, confirming that matter domination was a valid approximation.

Finally, we remark on a similar constraint that was recently published by Nakama, Suyama, Kohri, and Hiroshima in Ref.~\cite{nakama2018constraints} (hereafter NSKH).  Unlike previous UCMH works, this work did not employ the $\rho\propto r^{-9/4}$ density profile.  Instead, NSKH assumed that minihalos developed NFW density profiles, and like us, they constrained a delta-spiked power spectrum.  Thus, a comparison is warranted: despite assuming shallower density profiles, NSKH were able to derive comparable or stronger constraints on the integrated area $\mathcal{A}_0$ ($\mathcal{A}^2$ in their paper) of the power spectrum spike.

To model the NFW fitting parameters for their minihalos, NSKH assumed that halo concentrations $c=r_\mathrm{vir}/r_s$ grow at the rate $c\propto a^{1.575}$.  However, as we discussed in Sec.~\ref{sec:ucmh}, the inner profiles of dark matter halos tend to remain stable in time in the absence of disruptive events.  Under the same conditions, the virial radius $r_\mathrm{vir}$ in physical coordinates grows approximately as $a$.  If the concentration is growing much faster than $a$, then this implies that the physical scale radius is shrinking in time and the halo center is becoming denser.  Our simulations suggest that this is not the case: the concentrations of our minihalos grow as $c\propto a$.  The rate $c\propto a^{1.575}$ was drawn from a previous work \cite{zhao2009accurate} that simulated structure growth from scale-free power spectra.  With these power spectra, halo mergers are common, and these can cause halo physical virial radii $r_\mathrm{vir}$ to grow significantly faster than $a$.  This fact may explain the large concentration growth rate: it reflects rapid growth in virial radius $r_\mathrm{vir}$ rather than shrinkage in scale radius $r_s$.  Since halos that form from a spiked power spectrum do not experience these mergers, their concentrations grow more slowly.

Their assumption of a faster concentration growth rate likely explains why the constraints in NSKH are so strong.  The annihilation rates within such concentrated minihalos would be greatly enhanced.  However, NSKH also employed the diffuse gamma-ray flux from extragalactic sources, whereas we, for parity with BSA, assumed only Galactic sources.  This could contribute to the strength of their constraints: as we discuss above, the upper bound on the power spectrum is now highly sensitive to observational limits on the gamma-ray flux.

\section{Conclusion}\label{sec:conclusion}

Expanding on the results of Paper~I, we have shown that the minihalos that form due to a power spectrum enhancement do not develop the single-power-law ${\rho\propto r^{-9/4}}$ density profile even when they form by $z=1000$ from extremely rare ($6.8\sigma$) peaks.  Instead, they develop density profiles with inner power-law indices between $-3/2$ and $-1$, depending on the range of scales that are enhanced.  This finding contradicts the assumption made by previous UCMH work \cite{josan2010gamma,bringmann2012improved,li2012new,yang2013neutrino,yang2013dark,yang2014constraints,clark2015investigatingII,*clark2017erratumII,yang2017tau,aslanyan2016ultracompact,choi2017new}, throwing into question power spectrum constraints that have been derived from this theory.  However, we have also offered hope.  We constructed a new model based on our simulation results for minihalos that form from a spiked power spectrum, and we calculated a new power spectrum constraint in this model using Fermi-LAT constraints on gamma rays from WIMP annihilation.  The resulting upper bound on the primordial power spectrum is stronger than an equivalent constraint derived in the old UCMH picture.  It turns out that the drop in signal from each early-forming halo is more than compensated by the vast increase in the number of halos that contribute to the expected gamma-ray signal.

Our constraint is specialized to a power spectrum enhanced over a narrow range of scales.  Such spiked power spectra have motivations in inflationary phenomenology \cite{salopek1989designing,starobinsky1992,ivanov1994inflation,starobinsky1998beyond,chung2000probing,barnaby2009particle,barnaby2010features} and in nonstandard thermal histories of the Universe \cite{erickcek2011reheating,barenboim2014structure,fan2014nonthermal,erickcek2015dark}, but halos forming from these spectra have not been numerically studied prior to this work and Refs.~\cite{gosenca20173d,Delos2018ultracompact}.  Our model for halos forming from spiked power spectra predicts halo density profiles based on their formation time: the characteristic density is set by the background density at formation, while the characteristic scale is set by the spike scale at formation.  However, we developed this model based a single power spectrum.  We also neglected any scatter in the density profiles of halos forming at the same time.  In Paper~III, we will extend this model by quantifying its scatter and its applicability to different power spectra.

Inflationary phenomenology also includes less scale-localized power spectrum boosts such as steps or bends \cite{salopek1989designing,starobinsky1992,ivanov1994inflation,starobinsky1998beyond,joy2008new,silk1987double,polarski1992spectra,adams1997multiple}.  We have found that the halos forming from a stepped power spectrum develop the same density profiles as later-forming galaxy-scale halos.  Consequently, there is already a vast body of literature on modeling the density profiles of these halos (e.g.~\cite{navarro1996structure,navarro1997universal,jing2000density,eke2001power,bullock2001profiles,zhao2003mass,avila2005dependence,neto2007statistics,gao2008redshift,duffy2008dark,*duffy2011erratum,maccio2008concentration,zhao2009accurate,munoz2011redshift,klypin2011dark,giocoli2012formation,prada2012halo,ludlow2012dynamical,ludlow2013mass,bhattacharya2013dark,meneghetti2013reconciling,dutton2014cold,ludlow2014mass,diemer2015universal,correa2015accretion,okoli2015concentration,klypin2016multidark,angel2016dark}), and we expect that these results may be adapted toward constraining steps or bends in the power spectrum.

Our constraint also employed only gamma rays from WIMP annihilation in Galactic or near-Galactic sources.  We made this restriction in order to facilitate a direct comparison between an upper bound on the power spectrum derived in the shallower minihalo picture and an equivalent bound derived using the results of Ref.~\cite{bringmann2012improved}, which made the same restriction, used UCMHs forming at $z\geq 1000$, and assumed the $\rho\propto r^{-9/4}$ profile.  As a result, we have left open the possibility of immediately improving the power spectrum constraints by considering the diffuse annihilation signal from extragalactic minihalos as Refs.~\cite{yang2011constraints,nakama2018constraints} do.  We have also not explored the impact of the shallower density profiles on gravitational probes such as astrometric microlensing \cite{li2012new} and pulsar timing arrays \cite{clark2015investigatingI,*clark2016erratumI,clark2015investigatingII,*clark2017erratumII}.

Most pressingly, we neglected the influence of disruptive events on the minihalo abundance and their density profiles.  Minihalo-minihalo mergers are one such disruptive event.  They can be counted by means of Press-Schechter theory \cite{lacey1993merger}\footnote{However, the self-consistency of Press-Schechter merger rates is questioned in Ref.~\cite{benson2005self}; see also Ref.~\cite{neistein2008merger} for a counterpoint.} with a sharp $k$-space filter \cite{barkana2001constraints,benson2012dark,schneider2013halo}, but their physical impact, especially on minihalos with $\rho\propto r^{-3/2}$ inner profiles, is not yet well understood.  Reference~\cite{ogiya2016dynamical} simulated controlled halo mergers and observed that successive mergers cause the inner density profiles of these halos to relax toward shallower forms, an effect that we confirmed.  However, they also found that the merger product can have a higher central density than its progenitor halos.  Moreover, for highly unequal-mass mergers, a remnant of the smaller halo is expected to survive within the larger one \cite{berezinsky2008remnants}.

\begin{figure}[t]
	\centering
	\includegraphics[width=\columnwidth]{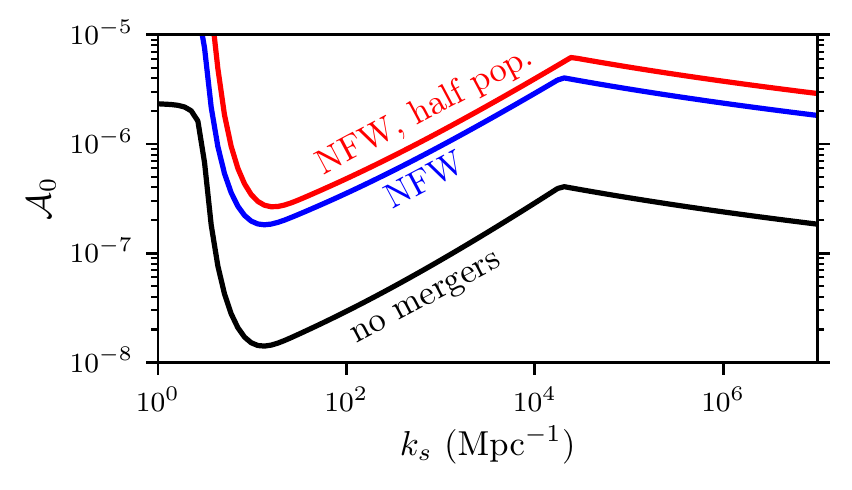}
	\caption{\label{fig:constraint_merge} Possible upper bounds on the integrated area $\mathcal{A}_0$ of a spike in the primordial curvature power spectrum centered at scale wave number $k_s$ when mergers are taken into account.  The black curve shows the bound with mergers neglected, which is the same constraint shown in Fig.~\ref{fig:constraint}.  The blue curve shows the bound if minihalos develop NFW profiles with the same scale parameters $r_s$ and $\rho_s$, while the red curve additionally halves the number of halos.}
\end{figure}

Figure~\ref{fig:constraint_merge} illustrates the possible impact of mergers on the minihalo-derived constraints on the primordial power spectrum.  If we naively assume that minihalos develop NFW profiles with the same scale parameters $r_s$ and $\rho_s$, then the shallower inner profiles weaken the power spectrum bound by roughly a factor of 10.  If mergers additionally halve the minihalo count, the constraint is weakened by another factor of $1.6$.  We suspect that this latter constraint, depicted as the red curve in Fig.~\ref{fig:constraint_merge}, represents a pessimistic estimate of how mergers may weaken the upper bound on the power spectrum.  First, we neglected the increased central densities that can result from mergers.  Second, we assumed that all halo profiles fully relax to NFW form, while Ref.~\cite{ogiya2016dynamical} showed that such relaxation is a gradual process occurring over multiple mergers.  Finally, mergers are relatively rare in spiked power spectra, and even if more than half of the minihalo population is ultimately destroyed by mergers (not even becoming subhalos), smaller halos, which contribute less to observational signals, are preferentially destroyed.  Hence, we expect that a careful accounting of mergers will produce a result between the black and red curves of Fig.~\ref{fig:constraint_merge}.

Disruption of minihalos can also occur by the tidal influence of larger galactic potentials or by high-speed encounters with objects, such as other substructure or stars, within these galactic potentials.  This topic has been studied in a number of previous works, such as Refs.~\cite{diemand2005earth,angus2007cold,berezinsky2006destruction,green2007mini,goerdt2007survival,zhao2007tidal,schneider2010impact}, and a recent overview of such disruptive processes can be found in Ref.~\cite{van2017disruption}.  It is also possible to bypass the issue of galactic disruption by only considering minihalos that have not accreted onto galactic halos, as Ref.~\cite{nakama2018constraints} does.

Our goal in this paper was to show that despite them not possessing the $\rho\propto r^{-9/4}$ density profiles that were previously assumed, minihalos are still able to yield competitive constraints on the primordial power spectrum.  The plethora of minihalos that now contribute to observational signals counteracts the loss of signal from the rarest of these halos.  This finding motivates their further study, and we have discussed avenues for future work.  With a better understanding of disruptive processes, minihalos can become strong and robust cosmological probes.

\begin{acknowledgments}
The simulations for this work were carried out on the KillDevil and Dogwood computing clusters at the University of North Carolina at Chapel Hill.  The authors would like to thank Erin Conn, Lucas deHart, Josh Horowitz, and Dayton Ellwanger for their valuable assistance in getting this project started on KillDevil.  The authors would also like to thank Peter Behroozi for supplying a beta version of \textsc{Rockstar}.  M.\,S.\,D. and A.\,L.\,E. were partially supported by NSF Grant No. PHY-1417446.  M.\,S.\,D. was also supported by the Bahnson Fund at the University of North Carolina at Chapel Hill.  A.\,P.\,B. contributed to this project while participating in the Computational Astronomy and Physics (CAP) Research Experiences for Undergraduates (REU) program funded by NSF Grant No. OAC-1156614 (PI S. Kannappan).
\end{acknowledgments}

\appendix

\section{Simulations prior to the matter-dominated era}\label{sec:rad}

For our numerical experiments, we employ a modified version of \textsc{Gadget-2} that includes a smooth radiation component, and we begin the simulations long before matter-radiation equality.  The simulation starting redshift of $z=8\times 10^6$ is necessary so that our enhanced fluctuations are still in the linear regime with amplitude $\delta\lesssim 0.1$.  However, the assignment of initial particle velocities is more complicated in this picture than during matter domination, and we use the Zel'dovich approximation to compute them in the following way.

We begin with the density contrast field $\delta(\vec q)$ as a function of the comoving grid coordinate $\vec q$.  We wish to convert this description into a comoving position field $\vec x(\vec q)$ and velocity field $\dot{ \vec x}(\vec q)$ treating $\vec q$ as a Lagrangian coordinate assigned to each particle.  The position calculation proceeds by writing
\begin{equation}\label{xq1}
\vec x(\vec q) = \vec q + \vec s(\vec q),
\end{equation}
where the displacement vector $\vec s$ is related to $\delta$ at linear order by
$
\nabla \cdot \vec s = -\delta
$.
If we assume $\vec s$ is irrotational, then $\vec s \propto \nabla \delta$, and the Fourier-transformed quantities are related by
\begin{equation}\label{sd1}
\vec s(\vec k) = \frac{i \vec k}{\vec k^2} \delta(\vec k).
\end{equation}
Equations (\ref{xq1}) and~(\ref{sd1}) determine the initial positions and are valid regardless of the composition of the universe.

We next turn to the velocity field $\dot{ \vec x}(\vec q)$ or its Fourier transform $\dot{ \vec x}(\vec k) = \dot{ \vec s}(\vec k)$.  Let $t_0$ be the time at which we are generating initial conditions, and write $\vec s$ as a function of time, using a new function $D(\vec k,t)$ to encode its time dependence:
\begin{equation}\label{st}
\vec s(\vec k,t) = D(\vec k,t) \vec s(\vec k).
\end{equation}
We define $\vec s(\vec k) \equiv \vec s(\vec k,t_0)$ so that $D(\vec k,t_0)\equiv 1$.  During matter domination, $D(\vec k,t) = a(t)/a(t_0)$ independent of $\vec k$, but radiation complicates the picture.  However, it is evident from Eq.~(\ref{sd1}) that $\vec s$ and $\delta$ evolve identically in time, implying $D(\vec k, t) = \delta(\vec k, t)/\delta(\vec k, t_0)$.  The initial velocity becomes
\begin{equation}\label{key}
\dot{\vec x}(\vec k,t_0) = \frac{\dot\delta(\vec k, t_0)}{\delta(\vec k, t_0)} \vec s(\vec k)
 = \left.\frac{\mathrm d \ln \delta(\vec k, t)}{\mathrm d t}\right|_{t=t_0} \vec s(\vec k),
\end{equation}
which is evaluated using Eq.~(\ref{evo}) with ${\mathrm d / \mathrm d t \equiv a H(a) \mathrm d / \mathrm d a}$.

To test the modified simulation code and initial conditions, we compare simulation results to linear theory.  We produce a matter power spectrum at $z=8\times 10^6$ using the procedure described in Sec.~\ref{sec:IC}, but we leave it unenhanced so that density contrasts near matter-radiation equality are well in the linear regime.  We draw initial conditions from this power spectrum in a (7.4 kpc)$^3$ periodic box and then evolve this box to $z=996$ using our modified version of \textsc{Gadget-2}\footnote{The fluctuations drawn from an unenhanced power spectrum have amplitude $\delta\sim 10^{-3}$ at $z=8\times10^6$, which results in extremely small particle accelerations.  To evade errors resulting from floating-point precision, we also set \textsc{Gadget-2} to use double-precision arithmetic.  All simulations in this paper employ this setting.}.  All simulation parameters are the same as those of the reference simulation described in Appendix~\ref{sec:conv}.  Figure~\ref{fig:rad_pkratio} shows the growth of the power spectrum during this simulation.  It matches the linear-theory prediction of Eq.~(\ref{evo}), including the scale-dependent growth.  Note that without the radiation component, the power spectrum would have instead grown by a factor of about $6\times 10^7$.

\begin{figure}[t]
	\centering
	\includegraphics[width=\columnwidth]{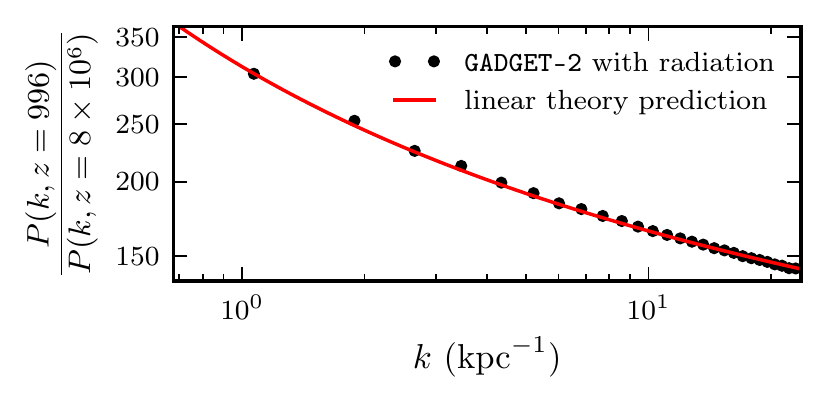}
	\caption{\label{fig:rad_pkratio} The power spectrum growth from $z=8\times 10^6$ to $z=996$: a comparison between \textsc{Gadget-2} with an added radiation component and linear theory.  In matter domination, the power spectrum would have instead grown by a factor of about $6\times 10^7$.}
\end{figure}

\begin{figure}[t]
	\centering
	\includegraphics[width=\columnwidth]{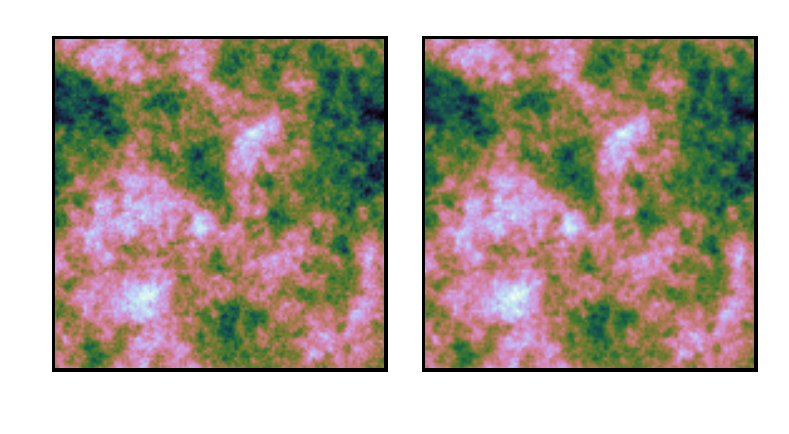}
	\caption{\label{fig:rad_field} Fractional overdensity fields evolved to $z=996$ from the same initial box at $z=8\times 10^6$.  The figure depicts a $(\SI{7.4}{kpc})^2\times 1.5 \si{kpc}$ slice.  The left and right panels show the results of linear theory and modified \textsc{Gadget-2} respectively.}
\end{figure}

As another demonstration, we also evolve the initial density field to $z=996$ using linear theory by applying the evolution specified by Eq.~(\ref{evo}) to the Fourier-transformed density field; we may then compare the resulting density field to the one evolved using \textsc{Gadget-2}.  In Fig.~\ref{fig:rad_field}, we plot a slice of the density field at $z=996$ evolved using both methods.  Our modified version of \textsc{Gadget-2} with initial conditions described above successfully reproduces the results of linear theory.

\section{Convergence testing and simulation parameters}\label{sec:conv}

\subsection{Simulation parameters}

We carry out a reference simulation run and five convergence-testing runs: one with improved force accuracy, one with improved integration accuracy, two with respectively increased and reduced softening scales, and one with higher particle count.  Our parameter choices for these runs are summarized in Table \ref{tab:convergence} and described below.  We refer the reader to Ref.~\cite{springel2005cosmological} for further detail on these parameters.

\paragraph{Particle count $N$}
In our reference run, we use $N=512^3$ particles arranged in a 3-dimensional grid.  In the high-particle-count run, we increase this to ${N=1024^3}$.  At the final redshift, the UCMH in the reference run has ${2.7\times 10^5}$ particles within its virial radius $r_\mathrm{vir}$ while the UCMH in the high-particle-count run has ${2.2\times 10^6}$ particles.

\paragraph{Particle mesh size $N_\mathrm{mesh}$}
This parameter describes the size of the particle mesh used for the long-range force calculation.  In our reference run, we set ${N_\mathrm{mesh} = N}$.  In the force-accuracy run, we set ${N_\mathrm{mesh} = 2^3N}$ to increase the accuracy of the long-range force calculation.

\paragraph{Short/long-range split $r_s$}
\textsc{Gadget-2} computes the long-range force using a particle mesh and the short-range force using an octree.  The parameter $r_s$ determines the splitting scale in units of mesh cells.  In our reference run, we set $r_s = 1.25$ mesh cells.  In the force-accuracy run, we increase this to $r_s = 2.5$ mesh cells, which effectively leaves the splitting scale unchanged since we also double the mesh frequency.

\paragraph{Short-range cutoff $r_\mathrm{cut}$}
The short-range force calculation (using an octree) is cut off beyond $r_\mathrm{cut}$ mesh cells.  In our reference run, we set $r_\mathrm{cut} = 4.5$ mesh cells.  In the force-accuracy run, we increase this to $r_\mathrm{cut} = 9.0$ mesh cells, which similarly leaves the cutoff scale unchanged since we also double the mesh frequency.

\begin{table}[t]
	\caption{\label{tab:convergence} Simulation parameters for convergence runs.  See the text for descriptions of the symbols.\vspace{1mm}}
	\begin{ruledtabular}
		\begin{tabular}{ccccccccc}
			
			label & $N$ & $\frac{N_\mathrm{mesh}}{N}$ & $r_s$ & $r_\mathrm{cut}$ & $\alpha$ & $\eta$ & $\mathrm{d}t_\mathrm{max}$ & $\frac{\epsilon}{\Delta r}$ \\
			\hline
			reference & $512^3$ & $1$ & 1.25 & 4.5 & 0.005 & 0.025 & 0.03 & 0.03 \\
			
			force acc. & $512^3$ & $2^3$ & 2.5 & 9.0 & 0.002 & 0.025 & 0.03 & 0.03 \\
			
			integration & $512^3$ & $1$ & 1.25 & 4.5 & 0.005 & 0.01 & 0.01 & 0.03 \\
			
			softening $\times 2$ & $512^3$ & $1$ & 1.25 & 4.5 & 0.005 & 0.025 & 0.03 & 0.06 \\
			
			softening $\times \frac{1}{3}$ & $512^3$ & $1$ & 1.25 & 4.5 & 0.005 & 0.025 & 0.03 & 0.01 \\
			
			N $\times 8$& $1024^3$ & $1$ & 1.25 & 4.5 & 0.005 & 0.025 & 0.03 & 0.03 \\
			
		\end{tabular}
	\end{ruledtabular}
\end{table}

\paragraph{Tree-force error parameter $\alpha$}
\textsc{Gadget-2}'s short-range (octree) force calculation only opens a tree node if the estimated force error from truncating it is less than $\alpha$ times the estimated total force.  In our reference run, we set $\alpha=0.005$.  In the force-accuracy run, we set $\alpha=0.002$ to increase the accuracy of the short-range force.

\paragraph{Adaptive time step parameter $\eta$}
\textsc{Gadget-2} uses individual adaptive time steps with an accuracy parameter $\eta$.  Roughly, the time step is set so that the maximum displacement due to a particle's acceleration over one time step is smaller than $\eta$ times the force-softening scale.  In the reference run, we set $\eta=0.025$, while in the integration-accuracy run, we set $\eta=0.01$ to reduce the particle time steps.

\paragraph{Maximum time step $\mathrm{d}t_\mathrm{max}$}
In order to avoid large integration errors at early redshifts when accelerations are small, \textsc{Gadget-2} imposes a maximum particle time step $\mathrm{d}t_\mathrm{max}$, which is expressed in units of the Hubble time (so it is actually $\mathrm d\ln a$).  In the reference run, we set $\mathrm{d}t_\mathrm{max} = 0.03$, while in the integration-accuracy run, we set $\mathrm{d}t_\mathrm{max} = 0.01$ to improve the integration accuracy at early times.

\paragraph{Force softening scale $\epsilon$}
\textsc{Gadget-2} softens the gravitational force based on the length parameter $\epsilon$, which we set to be a fraction $\epsilon/\Delta r$ of the initial interparticle spacing $\Delta r \equiv \mathrm{box\ size}/N^{1/3}$.  Note that the force becomes fully Newtonian at $2.8\epsilon$; $\epsilon$ itself is defined to be the minimum radius that can appear in the point-particle potential.  In the reference run, we set $\epsilon/\Delta r = 0.03$.  Larger softening lengths can minimize discreteness artifacts, but they also systematically bias the forces and prevent smaller scales from being resolved.  We perform two runs with altered softening: one with a larger softening scale $\epsilon/\Delta r = 0.06$ and one with a smaller softening scale $\epsilon/\Delta r = 0.01$.

\subsection{Procedure, results, and discussion}
We conduct convergence testing on the primary simulation run of Sec.~\ref{sec:ucmh}.  The density field is generated with $1024^3$ cells but reduced to $512^3$ cells for all but the high-particle-count run by averaging 8 neighboring cells.  Each convergence run is carried out as in Sec.~\ref{sec:ucmh} with only the simulation accuracy parameters changed.  We study here the spherically averaged density profile $\rho(r)$ of the UCMH at $z=100$.

We find that the density profile at each radius rapidly fluctuates between closely-spaced snapshots, and that these fluctuations differ between simulation runs with different parameters.  Moreover, we will see later that these fluctuations are a discreteness artifact (although they are not merely the Poisson noise in each radial bin, which is much smaller), so we wish to ignore them.  To do so, we obtain density profiles in 16 snapshots between $z=100$ and $z=99$ and average them.  We also compute the root-mean-squared variance between the snapshots as an estimate of the magnitude of this time variation.  Such a small time range was chosen so that we only smooth over rapid fluctuations and not over significant global evolution (such as growth).  Nevertheless, this time is short enough that it fails to average over fluctuations at large radii where the particle motion is much slower.  We may estimate the upper limit of the range over which rapid fluctuations are smoothed as the radius $r_\mathrm{lim}$ where 
\begin{equation}
\sigma_r(r_\mathrm{lim})\Delta t = r_\mathrm{lim},
\end{equation}
where $\sigma_r(r_\mathrm{lim})$ is the radial velocity dispersion at radius $r_\mathrm{lim}$ and $\Delta t$ is the time difference between $z=100$ and $z=99$.  $r_\mathrm{lim}$ is thus the radius at which particles are moving fast enough that their radial distance traveled over the averaging period is of the same order as their radial position.  For $r < r_\mathrm{lim}$, there can be no significant correlation between the positions of a particle at the beginning and at the end of the averaging period, so we expect to have averaged over these discreteness artifacts.

There is also an obvious lower limit to the range of radii over which we expect our results to be  representative, and that is where the radius is equal to $2.8\epsilon$, where $\epsilon$ is the gravitational softening length.  Below this radius, all forces are non-Newtonian, and the density profile will unphysically flatten out.

\begin{figure*}[t]
 \includegraphics[width=.49\linewidth]{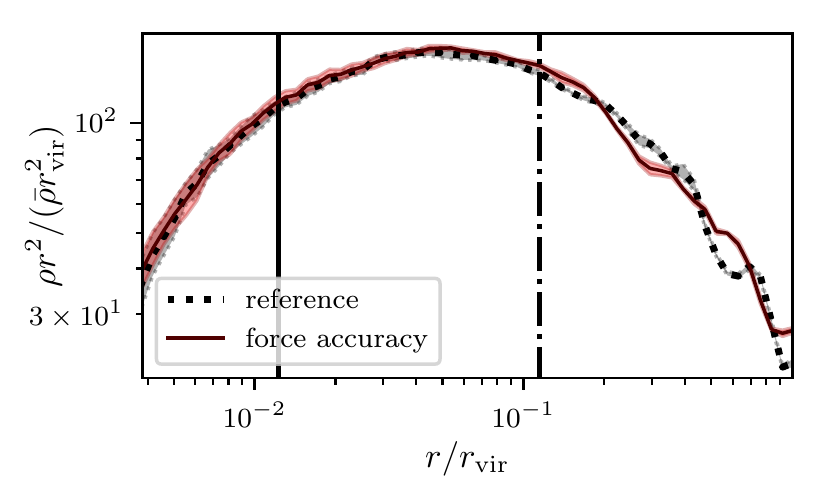} \hspace{-2mm}
 \includegraphics[width=.49\linewidth]{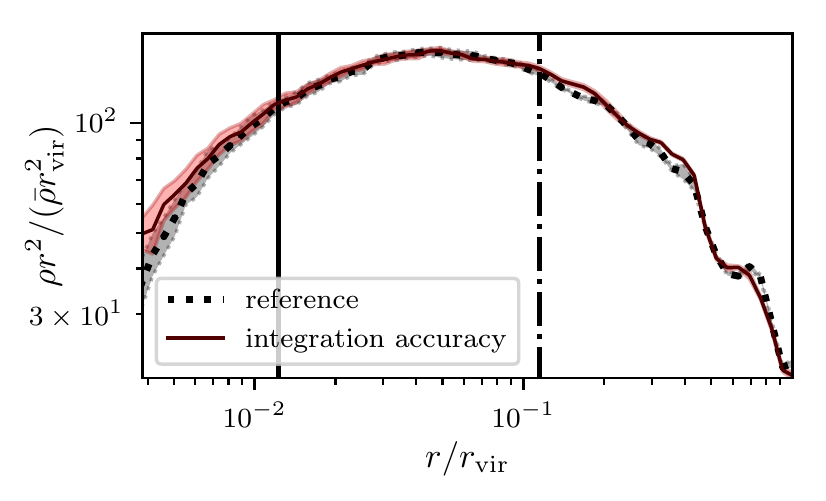} \\
 \vspace{-3mm}
 \includegraphics[width=.49\linewidth]{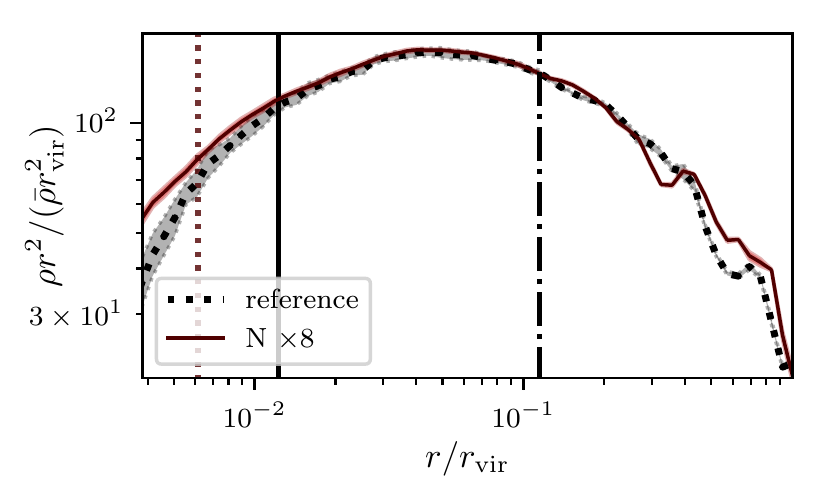} \hspace{-2mm}
 \includegraphics[width=.49\linewidth]{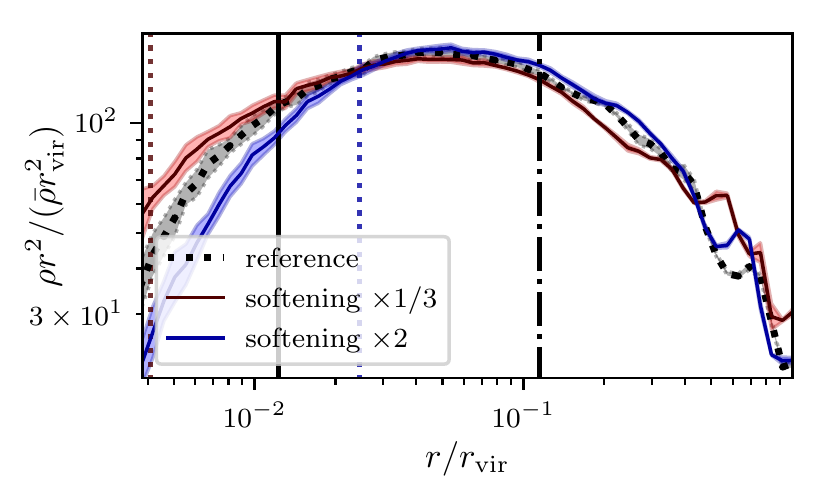}
\caption{\label{fig:conv} Radial density profiles averaged between $z=100$ and $z=99$; the shading indicates the root-mean-squared variance over this interval.  Clockwise from top left: force-accuracy, integration-accuracy, softening-length, and particle-resolution convergence comparisons.  The vertical solid (dotted) lines indicate $2.8\epsilon$ for the reference run (alternate runs), the range beyond which forces are exactly Newtonian.  The dot-dashed line indicates $r_\mathrm{lim}$, the radius beyond which fluctuations are likely not averaged (see text for details).  ($\bar\rho$ is the background matter density.)}
\end{figure*}

Our main results are shown in Fig.~\ref{fig:conv}.  The dotted lines indicate the density profile of the reference simulation run, with the gray shading representing the root-mean-squared variance in snapshots between ${z=100}$ and ${z=99}$.  The solid lines with colored shading depict alternate runs.

\subsubsection{Force accuracy}

In the force-accuracy run, we reduce the tree-force error parameter $\alpha$ to increase the accuracy of the short-range force and simultaneously double the resolution of the particle mesh in order to increase the accuracy of the long-range force.  We change the short/long range splitting parameters $r_s$ and $r_\mathrm{cut}$ only to keep the short/long range split the same, since those parameters are expressed in mesh cells.  The result is shown in Fig.~\ref{fig:conv}.  We see that for $r<r_\mathrm{lim}$, the density profile of the force-accuracy run matches well that of the reference run, suggesting that the reference force-accuracy parameters were sufficient.

\subsubsection{Integration accuracy}

In the integration-accuracy run, we reduce the integration time step in order to reduce error from the numerical integration of particle trajectories.  Specifically, we reduce the adaptive time step parameter $\eta$ as well as the maximum time step $\mathrm{d}t_\mathrm{max}$.  The first change should improve integration accuracy at late times when particles are experiencing large accelerations, and the second should improve integration accuracy at early times when accelerations are small.  The result is shown in Fig.~\ref{fig:conv}.  We see that for $2.8\epsilon <r<r_\mathrm{lim}$, the density profile of the integration-accuracy run matches that of the reference run, indicating that the reference integration-accuracy parameters were sufficient.

\subsubsection{Force softening}

The gravitational force softening length $\epsilon$ is a more difficult parameter to tune
\cite{dehnen2001towards,athanassoula2000optimal,power2003inner,
splinter1998fundamental,binney2002two,melott1997demonstrating}.  Forces are softened at short range to account for the way the numerical simulation uses discrete particles to represent a continuous mass distribution.  Thus, a larger $\epsilon$ reduces the influence of discreteness artifacts such as two-body collisions.  On the other hand, a larger $\epsilon$ also introduces a bias due to the forces being weaker than Newtonian, and a smaller $\epsilon$ can allow smaller scales to be probed.  Unlike the other parameters we consider, there is no clear direction of greater accuracy in $\epsilon$.  Ref.~\cite{power2003inner} suggests a minimum softening length $\epsilon_\mathrm{acc} \equiv r_\mathrm{vir}/\sqrt{N_\mathrm{vir}}$ ($N_\mathrm{vir}$ is the number of particles in the halo virial radius), which follows from the criterion that the maximum two-body acceleration caused by a close approach be smaller than the minimum mean-field acceleration within the system.  We adopt the choice $\epsilon=0.03\Delta r$, where $\Delta r$ is the initial interparticle spacing, in our reference run, which results in $\epsilon \simeq 2.3\epsilon_\mathrm{acc}$ at $z=100$.  We also execute two other simulation runs with $\epsilon$ at respectively a third and twice the reference value.

\begin{figure}[t]
  \centering
    \includegraphics[width=\columnwidth]{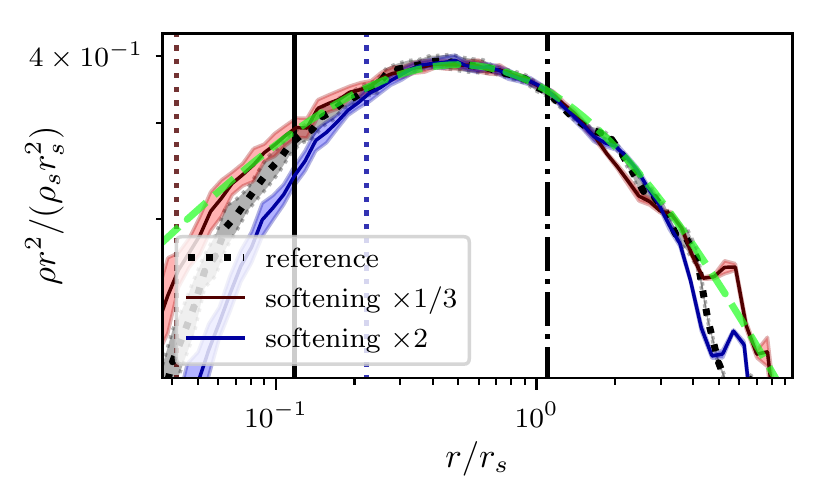}
    \caption{\label{fig:conv_soft2} The softening-length convergence test with $\rho$ and $r$ rescaled for each run to its Moore fitting parameters.  The thick dashed line shows the fit, which is the same for all runs by construction.  (Shading and vertical lines have the same meaning as in Fig.~\ref{fig:conv}.)}
\end{figure}

The results are shown in the bottom right of Fig.~\ref{fig:conv}.  The vertical solid line represents $2.8\epsilon$ for the reference run, while the two vertical dotted lines represent $2.8\epsilon$ for the runs with altered softening.  As one would expect, the density profile for each run flattens out for $r<2.8\epsilon$.  We hope to find convergence for $2.8\epsilon<r<r_\mathrm{lim}$ (recalling that each run has a different $\epsilon$), but while the density profiles in these ranges are close, there is a systematic flattening as $\epsilon$ becomes larger.  What has happened here may be elucidated by performing a Moore fit [Eq~(\ref{Moore})] to each run and rescaling $\rho$ and $r$ to the fitting parameters $\rho_s$ and $r_s$.  The rescaled plot is shown in Fig.~\ref{fig:conv_soft2}.  Evidently, each run can still be fit by the same form in the range $2.8\epsilon<r<r_\mathrm{lim}$ but has a different Moore concentration parameter $c\equiv r_\mathrm{vir}/r_s$: the reference run has $c=9.6$, the reduced softening run has $c=10.2$, and the increased softening run has $c=9.2$.  This discrepancy is likely the result of a force bias: the softening is causing a slight enlargement of the system.  Fortunately, the effect is small.  If we assume that either $r_s$ or $c$ is a linear function of $\epsilon$, then with a purely Newtonian force, we would have $c\simeq 10.4$, corresponding to an 8\% reduction in $r_s$ relative to the reference simulation.  Moreover, the softening length does not affect the major conclusion regarding the shape of the density profile: all runs fit the Moore form well for $r>2.8\epsilon$.

\subsubsection{Particle count}

We use $512^3=1.3\times 10^8$ particles in the reference run, which places $2.7\times 10^5$ particles within the virial radius of the UCMH at $z=100$.  To avoid strong discreteness effects, we must use enough particles that two-body collisions have negligible impact.  To estimate this, we consider the two-body relaxation timescale $t_\mathrm{relax}$, the timescale over which two-body encounters significantly alter a particle's energy.  For a region of radius $r$ about the halo center containing $N$ simulation particles with total mass $M$, $t_\mathrm{relax} = N/(8\ln N) t_\mathrm{cross}$ (e.g. \cite{binney1987galactic}) with $t_\mathrm{cross} \simeq r/\sqrt{GM/r}$.  The relaxation timescale should be much longer than the dynamical age of the halo, which is essentially the age of the universe at $z=100$.  We calculate $t_\mathrm{relax}$ for the UCMH in the reference simulation and find that even at the smallest relevant radius, $r=2.8\epsilon$, $t_\mathrm{relax}$ is 100 times the age of the universe at $z=100$.  This calculation suggests that the reference simulation contains enough particles that collisional artifacts are unimportant.

\begin{figure}[t]
  \centering
    \includegraphics[width=\columnwidth]{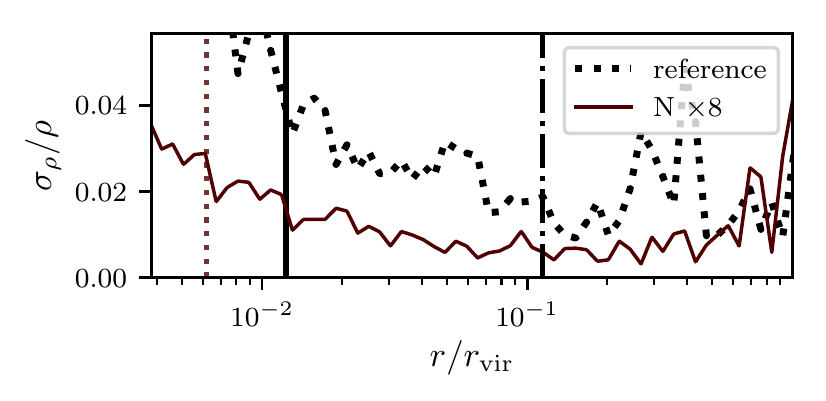}
    \caption{\label{fig:conv_highres_fluc} Root-mean-squared variance in the density profile across 16 snapshots between $z=100$ and $z=99$.  The reference and high-particle-count runs are compared.  (Vertical lines have the same meaning as in Fig.~\ref{fig:conv}.)}
\end{figure}

Nevertheless, Fig.~\ref{fig:conv} shows a comparison between the density profiles in the reference run and in a simulation run with 8 times as many particles.  As expected, the two density profiles match well in the range $2.8\epsilon<r<r_\mathrm{lim}$, implying that the simulation is converged with respect to particle count in these regions.  Moreover, the time-fluctuations in the density profile, measured as the root-mean-squared variance across snapshots between $z=100$ and $z=99$, are smaller in the high-particle-count simulation than in the reference simulation, a fact that we confirm in Fig.~\ref{fig:conv_highres_fluc}.  This observation confirms our claim that the fluctuations are a discreteness artifact.

\begin{figure}[t]
	\centering
	\includegraphics[width=\columnwidth]{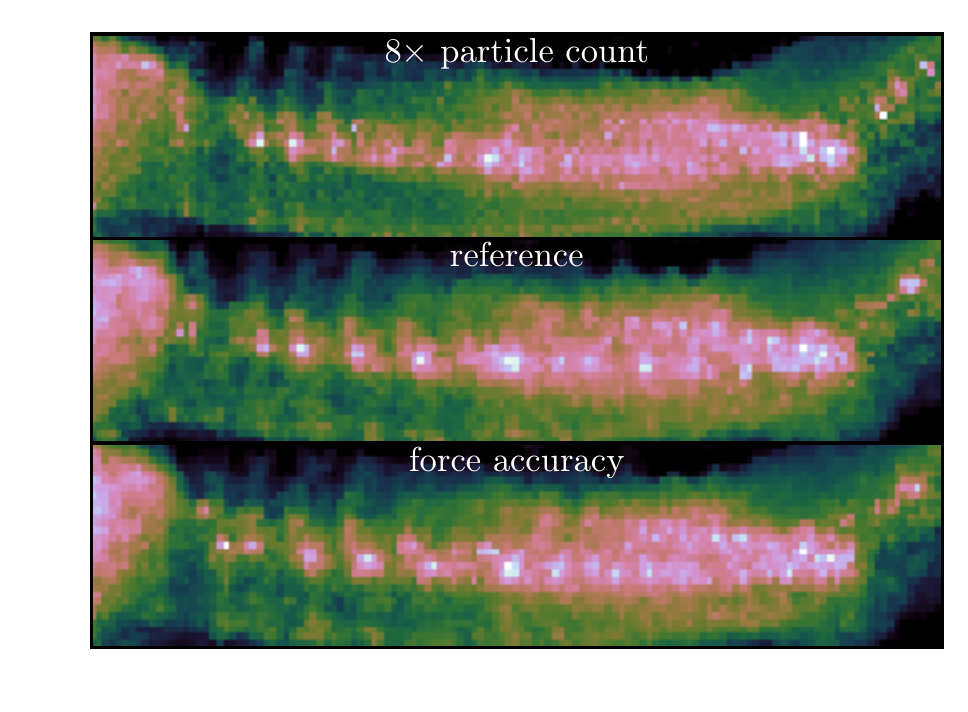}
	\caption{\label{fig:frag} The filament to the left of the UCMH (see Fig.~\ref{fig:halo}) for different simulation parameters.  This figure demonstrates the presence of artificial fragmentation: the filament fragments differently for different parameters.}
\end{figure}

These fluctuations may be related to the artificial fragmentation of filaments that occurs in simulations with a small-scale cutoff in the power spectrum \cite{angulo2013warm,lovell2014properties}.  We observe artificial fragmentation in our simulations using the spiked power spectrum, as evidenced in Fig.~\ref{fig:frag}, which shows one of the filaments connected to the UCMH for different simulation parameters.  Like the density fluctuations, the frequency and size of these fragments is correlated with the simulation particle resolution, while their positions vary with force-accuracy parameters.  The fluctuations in the density profile could be caused by the accretion of artificial fragments.

\subsection{The smallest resolved radius}

We have shown that the density profile of the UCMH in the reference version of the primary simulation run is converged with respect to simulation parameters at $z=100$ for radii $r$ between $2.8\epsilon= 0.012 r_\mathrm{vir}$ and ${r_\mathrm{lim}= 0.11 r_\mathrm{vir}}$.  This halo has $r_\mathrm{vir} = \SI{1.0e-3}{kpc}$ (in physical coordinates) at $z=100$, so the converged radius range is $\SI{1.2e-5}{kpc} < r < \SI{1.1e-4}{kpc}$. Moreover, we can regard the density profile above $r_\mathrm{lim}= \SI{1.1e-4}{kpc}$ as converged in its long-range behavior, with only the small-scale fluctuations being not converged.  Unfortunately, the lower limit of $\SI{1.2e-5}{kpc}$ imposed by the force softening is not sufficient to capture the asymptotic behavior in $\rho(r)$ at small $r$ (see Fig.~\ref{fig:ucmh}).

\begin{figure}[t]
  \centering
    \includegraphics[width=\columnwidth]{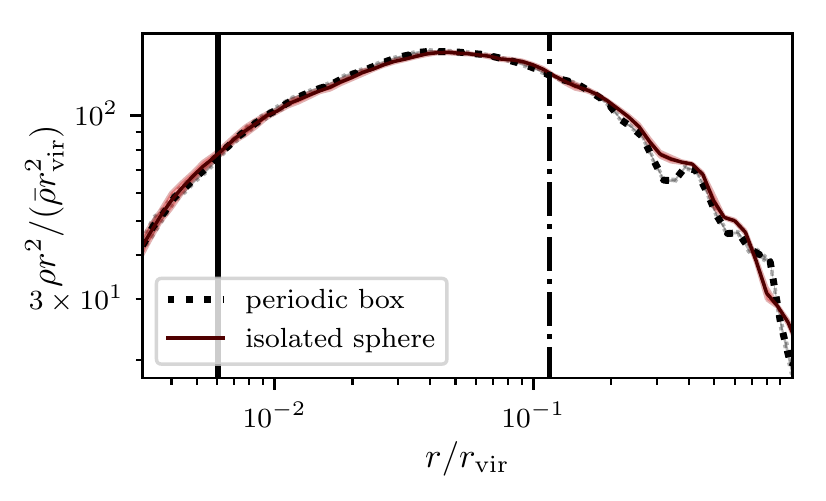}
    \caption{\label{fig:conv_reduce} A comparison between the UCMH simulated in a $(\SI{7.4}{kpc})^3$ periodic box with a halo from the same initial overdense region simulated in a sphere of radius $0.92\ \mathrm{kpc}$ with vacuum boundary conditions.  (Vertical lines have the same meaning as in Fig.~\ref{fig:conv}.)}
\end{figure}

We can double the resolution by employing the high-particle-count simulation run, but we would like to go still deeper into the halo.  While it is computationally challenging to simulate the full box with more than $1024^3$ particles, it is also unnecessary.  A common practice in N-body simulations is to resample the halo progenitor at higher particle resolution and embed this high-resolution region into the same periodic box.  Because the halos we consider are much more isolated than halos in a hierarchical growth picture, we need not even go this far: we can simply isolate a sphere around the halo progenitor and use vacuum boundary conditions.  Figure~\ref{fig:conv_reduce} shows the comparison between the periodic box with $1024^3$ particles and an otherwise identical simulation of a vacuum-bounded sphere of radius 0.92 kpc around the UCMH.  The spherical region is depicted in Fig.~\ref{fig:init}.  The sphere requires only $1/122$ as many particles for an identical result.  We therefore exploit this method to simulate the UCMH in the primary simulation at $64\times$ particle density and the other UCMHs at $8\times$ particle density relative to the reference simulation with $512^3$ particles.

However, this concordance does not hold at $z=50$, nor is it guaranteed to hold at $z=100$ in other boxes.  The longer the simulation run, the more likely the UCMH is to be influenced by structure that originated outside of the sphere.  Therefore, we restrict our use of the vacuum-bounded sphere to the primary simulation box (which we tested here) up to $z=100$ and to other initial boxes up to $z=400$.  For the UCMH in the primary simulation box, this brings the smallest resolved radius down to $\SI{3.0e-6}{kpc}$ (physical coordinates) at $z=100$, which is sufficient to resolve the beginning of the $\rho\propto r^{-3/2}$ asymptote.  This radius contains \num{3e4} particles at $z=100$.

Due to the stability of the density profile in time, we can probe still smaller radii by viewing the density profile at earlier times, as discussed in Section~\ref{sec:ucmh}.  By this method, we probe radii as small as \SI{9.0e-7}{kpc} in Fig.~\ref{fig:ucmh} using the density profile at $z=400$.  This radius contains 4000 particles at $z=400$ and is sufficient to demonstrate that the density profile shows no sign of leveling off toward a shallower power-law index than $3/2$.

\subsection{Summary of simulation choices}

We now summarize the simulation parameters we use to study the UCMHs in Sec.~\ref{sec:ucmh}.  Aside from the particle count $N$, all simulation runs use the reference parameters of Table~\ref{tab:convergence}. We employ two classes of simulation region: a comoving cube with periodic boundary conditions or an isolated comoving sphere with vacuum boundary conditions.  Table~\ref{tab:sims} shows the particle counts and simulation region sizes for all simulation runs.  The mass $m$ of the simulation particle is also shown for clarity.

\sisetup{round-mode=places,round-precision=1}
\begin{table*}[t]
	\caption{\label{tab:sims} The particle count $N$, particle mass $m$, simulation region size (side length or diameter), and ending redshift for each simulation used to produce the results in Sec.~\ref{sec:ucmh}.  The $\pi/6$ factor in the particle count comes from isolating a spherical region.  The number of particles $N_\mathrm{vir}$ within the UCMH is also shown at each redshift from which results are presented.  For the secondary simulations, an average figure is given.\vspace{1mm}}
	\begin{ruledtabular}
		\begin{tabular}{cccccccccc}
			
			description & $N$ & $m$ ($M_\odot$) & size (kpc) & region & end $z$ & $\left.N_\mathrm{vir}\right|_{z=400}$ & $\left.N_\mathrm{vir}\right|_{z=200}$ & $\left.N_\mathrm{vir}\right|_{z=100}$ & $\left.N_\mathrm{vir}\right|_{z=50}$ \\
			\hline\\
			
			primary  & $1024^3$       &$\num{1.5e-5}$& 7.4 & periodic & 100 & \num{549103} & \num{1252010} & \num{2242797} & ... \\
			
			primary, $z=50$ & $512^3$ &$\num{1.2e-4}$& 7.4 & periodic & 50 & ... & ... & ... & \num{431978} \\
			
			primary, vacuum b.c. & $\pi/6\times 512^3$ &$\num{1.8e-6}$& 1.85 & sphere & 100 & \num{4551792} & \num{10679480} & \num{18572202} & ...\\
			
			secondary & $512^3$      &$\num{1.2e-4}$& 7.4 & periodic & 50 & \num{71108.25} & \num{169121.5} & \num{284310.875} & \num{423624.5} \\
			
			secondary, vacuum b.c. & $\pi/6\times 256^3$ &$\num{1.5e-5}$& 1.85 & sphere & 400 & \num{583019.875} & ... & ... & ... \\
			
			step & $512^3$        &$\num{1.2e-4}$& 7.4 & periodic & 100 & ... & ... & \num{11631568} & ... \\
			
		\end{tabular}
	\end{ruledtabular}
\end{table*}

The primary simulation box is the same 7.4 kpc box we used for convergence testing.  The full box is simulated at 8 times the reference particle density ($1024^3$ particles) and a sphere around the main halo at 64 times the reference particle density up to $z=100$, and a third run simulates the full box at reference particle density up to $z=50$.  The density profile of the UCMH shown in Fig.~\ref{fig:ucmh} comes from the isolated sphere at $64\times$ reference particle density up to $z=100$ and from the full box at reference density at $z=50$.  Secondary simulations (Fig.~\ref{fig:ucmhs}) are executed at reference particle density up to $z=50$ and at $8\times$ reference density in an isolated sphere up to $z=400$.  The $N_\mathrm{vir}$ figures for the secondary simulations in Table~\ref{tab:sims} are average values and vary by up to 25\% between simulations of different UCMHs.  The density profiles of younger halos (Sec.~\ref{sec:later}) come from the full primary box at $8\times$ reference density at $z=100$.  Finally, for the stepped power spectrum, we simulated a full box with reference parameters to $z=100$.

All density profiles are averaged over 16 snapshots within $1/100$ of a Hubble time to suppress fluctuations.  Density profiles are binned logarithmically in intervals separated by a factor of 1.1 (corresponding to an interval of 0.041 in $\log_{10} r$), but we have checked that the results depend negligibly on this choice.

\begin{widetext}
\section{Constraining point-source abundance in the Milky Way}\label{sec:ptunif}
	
In Sec.~\ref{sec:constraints}, we defined $\mu(d) = 3M(d)/(4\pi d^3\bar\rho_0)$, where $M(d)$ is the dark matter mass contained within distance $d$ of Earth.  If we are at distance $r_0$ from the Milky Way center, then we may write
\begin{equation}\label{mu1}
\mu(d)=\frac{3}{2d^3\bar\rho_0}
\int_0^d s^2\mathrm d s
\int_{-1}^{1}\mathrm d x\ \max\!\left\{\rho_\mathrm{MW}\!\left(\sqrt{s^2+r_0^2-2x s r_0}\right),\bar\rho_0\right\},
\end{equation}
with $\rho_\mathrm{MW}(r)$ being the density profile of the Milky Way.  This expression approximates the extragalactic density field as a uniform background.  Taking the Milky Way to have an NFW profile with scale radius $r_\mathrm{S}$ and scale density $\rho_\mathrm{S}$ and the sun to be at distance $r_0$ from the center, the integral in Eq.~(\ref{mu1}) evaluates to
\begin{equation}\label{mu}
\mu(d)\simeq\frac{3 r_\mathrm{S}^3 \rho_\mathrm{S}}{2 r_0 \bar\rho_0 d^3}
\begin{cases}
(r_0+r_s)\ln\left(\frac{r_0+r_\mathrm{S}+d}{r_0+r_\mathrm{S}-d}\right)-2d
& d < r_0\\
2r_\mathrm{S}\mathrm{arctanh}\left(\frac{r_0}{d+r_\mathrm{S}}\right)+r_0\ln\left(\frac{(d+r_\mathrm{S})^2-r_0^2}{e^2r_\mathrm{S}^2}\right)
& r_0 < d < r_\mathrm{max}\\
2r_0\left[\frac{-r_\mathrm{max}}{r_\mathrm{max}+r_\mathrm{S}}+\ln\left(1+\frac{r_\mathrm{max}}{r_\mathrm{S}}\right)+\frac{\bar\rho_0}{3r_\mathrm{S}^3\rho_\mathrm{S}}(d^3-r_\mathrm{max}^3)\right]
& d > r_\mathrm{max}
\end{cases}
\end{equation}
where $r_\mathrm{max}$ is defined such that $\rho_\mathrm{MW}(r_\mathrm{max}) = \bar\rho_0$.  The first two cases in Eq.~(\ref{mu}) come from an exact evaluation of the integral in Eq.~(\ref{mu1}) for the Milky Way without a background, while the third case approximates $r_0\simeq 0$ to evaluate the extragalactic contribution.  Figure~\ref{fig:mu} shows a plot of $\mu(d)$.
\end{widetext}

\section{Constraining the power spectrum using the statistics of peaks}\label{sec:bbks}

In Sec.~\ref{sec:constraints}, we used the statistics of peaks as formulated in Ref.~\cite{bardeen1986statistics} (BBKS) to relate the differential halo number density by formation time, $\mathrm{d}n/\mathrm{d}a_c$, to the power spectrum $\mathcal{P}(k)$.  We describe that calculation in more detail here.  The differential number density of peaks according to their height $\nu=\delta/\sigma$ and steepness parameter $x$ is given in BBKS Eq.~(A14), where $\sigma$ is the root-mean-squared density variance\footnote{There is no smoothing filter, so $\sigma$ is the pointwise variance, ${\sigma^2=\int \frac{\mathrm{d}k}{k}\mathcal{P}(k)}$.} and $\delta$ is the peak density contrast.  For the spiked power spectrum given by Eq.~(\ref{pkdelta}), $\sigma = \mathcal{A}^{1/2} a$.  Moreover, the spectral parameter $\gamma$ [BBKS Eq.~(4.6)] has value $\gamma=1$, causing the $x$-exponential in BBKS Eq.~(A14) to become a delta-function.  We obtain the halo abundance by integrating this differential number density over $x$ and $\nu$ subject to the collapse requirement $\nu\sigma > \delta_c$, with result
\begin{equation}\label{n_bbks}
n = \frac{k_s^3}{(2\pi)^2 3^{3/2}} \int_{\delta_c/(\mathcal{A}^{1/2} a)}^\infty e^{-\nu^2/2}f(\nu) \mathrm d \nu
\end{equation}
where the function $f(\nu)$ is defined by BBKS Eq.~(A15).

We seek $\mathrm d n/\mathrm d a$, which is obtained by differentiation as
\begin{equation}
\label{dnda_bbks}
\frac{\mathrm{d}n}{\mathrm{d}a} = \frac{k_s^3}{a}\ h\!\left(\frac{\delta_c}{\mathcal{A}^{1/2}a}\right)
\end{equation}
with
\begin{equation}\label{dnda_func}
h(\nu)\equiv \frac{\nu}{(2\pi)^2 3^{3/2}} e^{-\nu^2/2} f(\nu).
\end{equation}
As we noted in Sec.~\ref{sec:powerspectrum}, the number density $n$ increases monotonically due to halo formation alone.  Thus, Eq.~(\ref{dnda_bbks}) gives us precisely $\mathrm d n/\mathrm d a_c$, which we can combine with Eq. (\ref{n_pt}) or~(\ref{n_diff}) to constrain the integrated area $\mathcal{A}$ of the power spectrum.

\begin{figure}[t]
	\centering
	\includegraphics[width=\columnwidth]{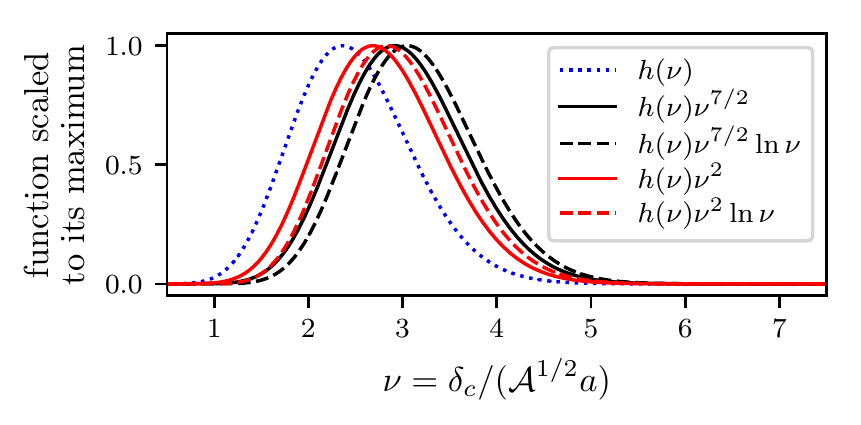}
	\caption{\label{fig:dnda_h} Plots of the function $h$ defined in Eq.~(\ref{dnda_func}) along with the integrands in Eq.~(\ref{integrals_def}).  Each function is scaled to its maximum value to emphasize the support of these functions.  The black curves are relevant to point-source constraints, while the red curves determine diffuse constraints.  This figure shows that constraints are set primarily by peaks between $2\sigma$ and $4\sigma$.}
\end{figure}

For the general case of point sources with nonconstant $\mu(d)$, we have little choice but to numerically invert the integral in Eq.~(\ref{n_pt}) to obtain an upper bound on $\mathcal{A}$ as a function of $k_s$.  However, in a limiting case where $\mu$ is constant, or to derive a bound from the diffuse flux, we can fully extract the $\mathcal{A}$-dependence from the integral.
In these cases, we have an integral of the form
\begin{align}\label{constraint_calc}
&\int_0^1 \mathrm{d}a_c \frac{\mathrm{d}n}{\mathrm{d}a_c} L^p(a_c)
= k_s^3\int_0^\infty \frac{\mathrm{d}\nu}{\nu}h(\nu) L^p\left(\frac{\delta_c}{\nu\mathcal{A}^{1/2}}\right)
\nonumber\\
&= \frac{A^p \mathcal{A}^{3p/2}}{\delta_c^{3p}k_s^{3p-3}}\int_0^\infty h(\nu)\nu^{3p-1} \left(\ln\frac{B\delta_c}{\nu\mathcal{A}^{1/2}}\right)^p\mathrm{d}\nu
\nonumber\\
&= \frac{A^p \mathcal{A}^{3p/2}}{\delta_c^{3p}k_s^{3p-3}}
\left[
\left(\ln\frac{B\delta_c}{\mathcal{A}^{1/2}}\right)^p I_p
-\left(\ln\frac{B\delta_c}{\mathcal{A}^{1/2}}\right)^{p-1}J_p 
\right],
\end{align}
with $p=3/2$ or $p=1$ for point sources or the diffuse flux respectively, and where $I_p$ and $J_p$ are defined as
\begin{align}\label{integrals_def}
I_p \equiv \int_0^\infty h(\nu)\nu^{3p-1}\mathrm{d}\nu,
\ 
J_p \equiv p\int_0^\infty h(\nu)\nu^{3p-1}\ln \nu\ \mathrm{d}\nu .
\end{align}
In the first line, we use Eq.~(\ref{dnda_bbks}) with $\nu=\delta_c/(\mathcal{A}^{1/2}a)$.  We saw in Sec.~\ref{sec:constraint_discussion} that the integral in Eq.~(\ref{constraint_calc}) is dominated by minihalos forming at $z\gtrsim 20$, so we exploit the negligible contribution to this integral of minihalos forming at $a>1$ to extend the lower limit of the integral on the right-hand side to $\nu=0$.  In the second line, we specialize to the Moore density profile using Eq.~(\ref{Moorelumshort}).  In the last line, we take advantage of the limited support of $h(\nu)$ to claim that $\ln \nu \ll \ln \left( B \delta_c/\mathcal{A}^{1/2} \right)$ so that we can use a binomial expansion (and this is exact for $p=1$).  Note that the integrands in Eq.~(\ref{integrals_def}) tell us the range of peaks that are relevant to constraints: as we see in Fig.~\ref{fig:dnda_h}, most of their support lies in peaks between roughly $2\sigma$ and $4\sigma$.

For point sources with $\mu\simeq\mathrm{const}$, we now have 
\begin{align}
\mathcal{A}&\left(\ln\frac{B\delta_c}{\mathcal{A}^{1/2}}\right)^{2/9}\left[\left(\ln\frac{B\delta_c}{\mathcal{A}^{1/2}}\right)I_{3/2}-J_{3/2}\right]^{4/9}
\nonumber\\
\tag{\ref{pt_muconst}}
&\leq
\left(\frac{-3\sqrt{4\pi}\ln(1-y/x)}{\mu}\right)^{4/9}\left(\frac{\delta_c^3 k_s \mathcal F_\mathrm{min}}{A}\right)^{2/3},
\end{align}
while for diffuse sources, we have
\begin{align}
\tag{\ref{diffuse}}
\mathcal{A}\left[\left(\ln\frac{B\delta_c}{\mathcal{A}^{1/2}}\right)I_1-J_1\right]^{2/3}
\leq
\left(\frac{4\pi\delta_c^3}{K(\theta)A}\frac{\mathrm d \mathcal F}{\mathrm d \Omega}\right)^{2/3}.
\end{align}
These expressions follow from Eqs. (\ref{n_pt}), (\ref{n_diff}), and~(\ref{constraint_calc}).  The numbers $I_{3/2}$, $J_{3/2}$, $I_1$, and $J_1$ have approximate values
\begin{align}\label{integrals}
I_{3/2} = 0.228, \ &\ J_{3/2} = 0.370
\nonumber\\
I_1 = 0.0477, \ &\ J_1 = 0.0478
\end{align}
Equations (\ref{pt_muconst}) and~(\ref{diffuse}) are now algebraic equations for the upper bounds on $\mathcal A$.

The constraint on $\mathcal{A}$ is not our final goal, but it is close.  $\mathcal{A}a^2$ is the power associated with the spike in the (linear) matter power spectrum during matter domination.  We seek instead the power $\mathcal{A}_{0}$ associated with the spike in the primordial curvature power spectrum.  These quantities are related by a transfer function such as \cite{weinberg2008cosmology}
\begin{equation}\label{deltaMD}
\delta(k,a) = \frac{2}{5}\frac{k^2}{\Omega_m H_0^2}\zeta(k)\ \mathcal{T}\!\left(\frac{\sqrt{\Omega_r}k}{H_0 \Omega_m}\right)a,
\end{equation}
which relates the matter density contrast $\delta$ during matter domination to the primordial curvature fluctuation $\zeta$.  Here 
\begin{equation}\label{deltaMD_transfer}
\mathcal{T}(x) = \frac{45}{2x^2}\left(-\frac{7}{2}+\gamma_E+\ln\!\left(\frac{4x}{\sqrt 3}\right)\right)
\end{equation}
is a dimensionless transfer function that is valid at $x \gg 1$ or $k\gg 10^{-2}\si{Mpc^{-1}}$ ($\gamma_E\simeq 0.577$ is the Euler-Mascheroni constant).  Hence, by squaring Eq.~(\ref{deltaMD}),
\begin{equation}\label{Pzeta}
\mathcal{A}_{0} = \frac{\left(\Omega_r/\Omega_m\right)^2\mathcal{A}}{81\left[-\frac{7}{2}+\gamma_E+\ln\!\left(\frac{4\sqrt{\Omega_r}k_s}{\sqrt 3 H_0 \Omega_m}\right)\right]^2}
\end{equation}
yields the desired constraint on the primordial power spectrum.

\section{The UCMH constraint on a spiked power spectrum}\label{sec:bsa}

Bringmann, Scott, and Akrami \cite{bringmann2012improved} (BSA) calculated an upper bound on the number density of UCMHs (as a function of scale wave number $k$) using the $\rho\propto r^{-9/4}$ density profile from Ref.~\cite{ricotti2009new}.  They then converted this constraint into an upper bound on the primordial power spectrum under the assumption of local scale invariance.  We followed the calculation in BSA as closely as possible in Sec.~\ref{sec:constraints} so as to facilitate a direct comparison in constraining strength between our new minihalo model and the old UCMH model.  However, because a spiked power spectrum does not exhibit local scale invariance, we must return to the UCMH abundance constraint in BSA and convert it into a constraint on the delta-spiked power spectrum given by Eq.~(\ref{pkdelta}).  We show that calculation here.

BSA Fig.~1 shows their constraint on the fraction $f$ of matter contained in UCMHs.  This fraction is readily converted into a number density $n=f\rho_m/M_\mathrm{UCMH}$, where BSA took
\begin{equation}\label{bsa_m}
M_\mathrm{UCMH}=\num{4e13}\left(\frac{k}{\mathrm{Mpc}^{-1}}\right)^{-3}M_\odot.
\end{equation}
Here we have employed $R=1/k$, where $R$ is the comoving radius of the precursor overdense region; this is the same relation BSA used.  The UCMHs are taken to follow the dark matter distribution, so $n$ and $\rho_m$ are the comoving background UCMH number density and matter density respectively.

This procedure has given us a constraint $n$ on the comoving number density of halos of scale wavenumber $k$ forming at $z\gtrsim 1000$.  In the delta-spiked power spectrum given by Eq.~(\ref{pkdelta}), all halos form from fluctuations with wavenumber $k_s$, the wavenumber of the spike, so we will take $k=k_s$ in Eq.~(\ref{bsa_m}).  We can then use Eq.~(\ref{n_bbks}) with $a=10^{-3}$ to convert this upper bound on the abundance of halos that form at $z\geq 1000$ into a constraint on the primordial power spectrum, which is shown in Fig.~\ref{fig:constraint}.
\newpage
\bibliography{UCMHdensityreferences}

\end{document}